\documentclass[pra,aps,twocolumn,groupedaddress,nofootinbib]{revtex4}
\usepackage{graphicx}
\usepackage[nointegrals]{wasysym} 
\usepackage[export]{adjustbox}
\usepackage{amsmath,amsfonts,amssymb,latexsym}
\usepackage{hhline}
\usepackage{bm}
\usepackage{verbatim}
\usepackage{enumitem}
\usepackage{mathrsfs}
\usepackage{slashed}
\usepackage{empheq}

\newcommand{\Tr}{\text{Tr}}

\newcommand{\wh}{\widehat}
\newcommand{\p}{\partial}

\newcommand{\lan}{\langle}
\newcommand{\ran}{\rangle}

\newcommand{\unit}{\mathbf{1}}

\newcommand{\da}{{\dagger}}
\newcommand{\doa}{\downarrow}
\newcommand{\upa}{\uparrow}
\newcommand{\ob}[1]{\mkern 1.5mu\overline{\mkern-1.5mu#1\mkern-1.5mu}\mkern 1.5mu}

\newcommand{\ra}{\rightarrow}

\newcommand{\wt}{\widetilde}

\newcommand{\uvx}{{\mathbf{\hat x}}}
\newcommand{\uvy}{{\mathbf{\hat y}}}
\newcommand{\uvz}{{\mathbf{\hat z}}}

\renewcommand{\(}{\left(}
\renewcommand{\)}{\right)}
\renewcommand{\[}{\left[}
\renewcommand{\]}{\right]}
\newcommand{\mt}{\mapsto}

\newcommand{\tp}{\otimes}

\newcommand{\twp}{{2\pi}}

\newcommand{\D}{\nabla}

\newcommand\bpm            {\begin{pmatrix}}
	\newcommand\epm           {\end{pmatrix}}

\newcommand{\ms}{\medskip}

\def\app#1#2{%
	\mathrel{%
		\setbox0=\hbox{$#1\sim$}%
		\setbox2=\hbox{%
			\rlap{\hbox{$#1\propto$}}%
			\lower1.1\ht0\box0%
		}%
		\raise0.25\ht2\box2%
	}%
}

\newcommand{\sech}{{\rm sech}}

\newcommand{\vp}{\varphi}
\newcommand{\vpi}{\varpi}
\newcommand{\vk}{\varkappa} 
\newcommand{\vup}{\varUpsilon}
 
\newcommand{\vs}{\varsigma}

\newcommand{\inv}{^{-1}}

\newcommand{\ope}\odot

\usepackage{manfnt}

\renewcommand{\prl}{{\, \parallel \,}}
\newcommand{\bi}{\begin{itemize}}
\newcommand{\ei}{\end{itemize}}

\usepackage{amsthm}

\newcommand{\bproof}{\begin{proof}}
\newcommand{\eproof}{\end{proof}}
\newtheorem{theorem}{Theorem}
\newtheorem{corollary}{Corollary}
\newtheorem{proposition}{Proposition}
\theoremstyle{definition}
\newtheorem{definition}{Definition}
\theoremstyle{definition}

\newtheorem{example}{Example}
\newcommand\bd            {\begin{definition}}
\newcommand\ed            {\end{definition}}
\newcommand\bt            {\begin{theorem}}
\newcommand\et            {\end{theorem}}
\newcommand\bpro		  {\begin{proposition}}
\newcommand\epro 		  {\end{proposition}}
\newcommand\bcor			{\begin{corollary}}
\newcommand\ecor		{\end{corollary}}
\newcommand\bex			{\begin{example}}
\newcommand\eex			{\end{example}}
\newcommand\bpr			  {\begin{proof}}
\newcommand\epr 		  {\end{proof}}

\newcommand\be            {\begin{equation}}
\newcommand\ee            {\end{equation}}
\newcommand\ba            {\begin{aligned}}
\newcommand\ea            {\end{aligned}}
\newcommand\bea{\begin{equation}\begin{aligned}}
\newcommand\eea{\end{aligned}\end{equation}}

 \usepackage{hyperref} 
 \hypersetup{final}

 \newcommand{\sss}{\subsubsection}
 \renewcommand{\ss}{\subsection}
 
 \renewcommand{\a}{\alpha}
 \renewcommand{\b}{\beta}
 \renewcommand{\d}{\delta}
 \newcommand{\De}{\Delta}
 \newcommand{\g}{\gamma}
 \newcommand{\G}{\Gamma}
 \newcommand{\s}{\sigma}

 \newcommand{\ep}{\varepsilon} 
 \renewcommand{\l}{\lambda}
 \renewcommand{\L}{\Lambda}
 \renewcommand{\t}{\theta}
 
 \renewcommand{\o}{\omega}

 \renewcommand{\r}{\rho}

 \newcommand{\z}{\zeta}

 \newcommand{\bfsig}{{\boldsymbol{\sigma}}}

 \newcommand{\bfxi}{{\boldsymbol{\xi}}}
 \newcommand{\bfeta}{{\boldsymbol{\eta}}}
 
 \newcommand{\bfA}{\mathbf{A}}
 \newcommand{\bfB}{\mathbf{B}}

 \newcommand{\bfI}{\mathbf{I}}

 \newcommand{\bfM}{\mathbf{M}}

 \newcommand{\bfa}{\mathbf{a}}
 \newcommand{\bfb}{\mathbf{b}}
 \newcommand{\bfc}{\mathbf{c}}
 \newcommand{\bfd}{\mathbf{d}}

 \newcommand{\bfk}{\mathbf{k}}
 
 \newcommand{\bfn}{\mathbf{n}}

 \newcommand{\bfq}{\mathbf{q}}
 \newcommand{\bfr}{\mathbf{r}}
 \newcommand{\bfs}{\mathbf{s}}

 \newcommand{\bfx}{\mathbf{x}}

 \newcommand{\zt}{\mathbb{Z}_2}

 \newcommand{\zz}{\mathbb{Z}}

 \newcommand{\mck}{\mathcal{K}}

 \newcommand{\mce}{\mathcal{E}}

 \newcommand{\mcl}{\mathcal{L}}
 \newcommand{\mcg}{\mathcal{G}}
 
 \newcommand{\mct}{\mathcal{T}}

 \newcommand{\mcp}{\mathcal{P}}

 \newcommand{\mcm}{\mathcal{M}}

 \newcommand{\sfg}{\mathsf{g}}

 \usepackage[mathscr]{eucal} 

 \newcommand{\sch}{\mathscr{H}}

\newcommand{\qq}{\qquad}

\usepackage{braket}

\newcommand{\kb}[1]{\ket{#1}\bra{#1}}

\renewcommand{\k}{\ket}

\usepackage{dcolumn}

\newcommand{\isum}{\sum_{\o_n}\int_\bfk}

\newcommand{\ton}{T_{\rm onset}}
\newcommand{\tb}{T_{BKT}}

\newcommand{\bre}{B_{\rm re-entrant}} 
\usepackage{xcolor}
\newcommand{\cpl}{V}
\newcommand{\duu}{d_{+}}
\newcommand{\ddd}{d_{-}}
\newcommand{\dud}{d_{z}}

\newcommand{\gminus}{\sfg}
\newcommand{\gplus}{\sfg}

\usepackage[caption=false]{subfig}
\begin{document}

	\title{Re-entrant superconductivity through a quantum Lifshitz transition in twisted trilayer graphene}
	
	\author{Ethan Lake}
	\author{T. Senthil}
	\affiliation{Department of Physics, Massachusetts Institute of Technology, Cambridge, MA, 02139}
	\begin{abstract}
	A series of recent experiments have demonstrated robust superconductivity in magic-angle twisted trilayer graphene (TTG). In particular, a recent work by Cao et al. (arxiv:2103.12083) studies the behavior of the superconductor in an in-plane magnetic field and an out-of-plane displacement field, finding that the superconductor is unlikely to have purely spin-singlet pairing. This work also finds that at high magnetic fields and a smaller range of dopings and displacement fields, it undergoes a transition to a distinct field-induced superconducting state. Inspired by these results, we develop an understanding of superconductivity in TTG using a combination of phenomenological reasoning and microscopic theory. We describe the role that that an in-plane field plays in TTG, and use this understanding to argue that the re-entrant transition may be associated with a quantum Lifshitz phase transition, with the high-field phase possessing finite-momentum pairing. We argue that the superconductor is likely to involve a superposition of singlet and triplet pairing, and describe the structure of the normal state. We also draw lessons for twisted bilayer graphene (TBG), and explain the differences in the phenomenology with TTG despite their close microscopic relationship. We propose that a singlet-triplet superposition is realized in the TBG superconductor as well, and that the $\nu = -2$ correlated insulator may be a time reversal protected $\mathbb{Z}_2$ topological insulator obtained through spontaneous spin symmetry breaking.
		
	\end{abstract}
	
	\maketitle
	
	\section{Introduction \label{sec:intro}} 
	
	A remarkable series of developments have brought two-dimensional graphene-based Moire materials to the spotlight of condensed matter physics. Many of these systems host a fascinating array of strongly correlated insulating and magnetic states \cite{cao2018correlated,chen2019evidence,sharpe2019emergent,chen2020tunable,serlin2020intrinsic,zondiner2020cascade,saito2020isospin,saito2021hofstadter,wu2021chern,chen2020electrically,cao2020nematicity,rozen2020entropic,zhou2021half,balents2020superconductivity,andrei2021marvels}, and twisted bilayer graphene (TBG) in particular displays robust unconventional superconductivity \cite{liu2020tunable,lu2019superconductors,yankowitz2019tuning,cao2018unconventional,arora2020superconductivity}. 
	Recently, another Moire system, twisted trilayer graphene (TTG), was also predicted and subsequently found to exhibit robust superconductivity \cite{khalaf2019magic,li2019electronic,hao2021electric,park2021tunable}. A very recent work \cite{inplane_fields} studies superconductivity in near-magic angle TTG in the presence of an in-plane magnetic field $\bfB$ and an out-of-plane electric displacement field $D$, and finds some strking phenomena as we review below. 
		In the present paper  we will synthesize the existing experimental results into a coherent theoretical picture of the superconductivity in TTG. We will  explain the differences with the closely related twisted bilayer graphene (TBG), and will show how comparing phenomena in TTG and TBG can give us important clues to understand both systems. 
	
	We begin with a discussion of some of the salient experimental results on superconductivity in near magic-angle TTG \cite{hao2021electric,park2021tunable,inplane_fields}. Our focus will be on the robust superconductivity  found for moire lattice fillings in the range $-3 < \nu < -2$. 
		\begin{enumerate} 
			\item
			The `normal' state out of which the superconductivity emerges has a carrier density that is set by the deviation from $\nu = -2$. This normal state  has a Landau fan degeneracy of $2$, rather than the $4$-fold degeneracy naively expected on the basis of the available flavor degrees of freedom (spin and valley). Furthermore the strength of the superconductivity does not seem to correlate with the single particle density of states inferred from indirect measurements. 
			
			\item
			The superconductivity occurs even without the external perturbations of the perpendicular displacement field $D$ or the in-plane magnetic field $B_\parallel$. Weak $D$ and $B_\parallel$ have contrasting effects on the  superconductivity - the former enhances it while the latter suppresses it. Strikingly  Ref. \cite{inplane_fields} finds that the superconducting state in a range of fillings survives in $B_\parallel$ fields far in excess of the Pauli limit, suggesting that the pairing is not spin-singlet in nature. This behavior is also notably different from  that seen in TBG, where the superconductivity is suppressed by  a $B_\parallel$ roughly at the Pauli limit. 
			
			\item
			
			Near optimal doping and displacement field strength, the superconductivity is seen to exhibit a ``re-entrant'' phenomenon at large magnetic fields, with a phase transition that separates two distinct superconductors occurring at a field $B_{\rm re-entrant} \sim 8$ T. Interestingly, while the temperature $T_{BKT}$ at which superconductivity occurs goes to zero at $B_{\rm re-entrant}$, the temperature $T_{\rm onset}$ at which the resistance reaches a fixed fraction of its normal-state value is smooth across the transition, and decreases monotonically with $B$. 
			
		\end{enumerate}
		
		In what follows,  by combining the input from experiments with a Ginzburg-Landau analysis, together with the insights from simple microscopic theory, we will provide an understanding of these phenomena, and make predictions for future experiments. We will be naturally led to the conclusion that the pairing in the superconductor is neither spin singlet {\it nor} spin triplet, but rather a linear combination of the two.  We examine the role that the $\bfB$ and $D$ fields play in the physics of TTG, and highlight differences with TBG. This will enable an explanation of the striking differences in the stability of the superconductor to in-plane fields in the two systems. We  propose  that interpreting the re-entrant transition as arising from a Lifshitz critical point, where one or both components of the superfluid stiffness vanish and where the pairing in the high-field phase occurs at finite momentum, is particularly natural in light of the existing experimental data.  
	
	Along the way we also discuss the nature of the normal state near filling $\nu \approx - 2$. We propose that at $\nu = -2$ the ground state is a time reversal protected topological insulator (obtained through spontaneous spin symmetry breaking) coexisting with the massless Dirac fermions corresponding to a decoupled graphene sector. 
	
	At $D=0,B=0$, the physics of TTG is very closely related to that of TBG \cite{khalaf2019magic}, as will be reviewed shortly. This leads us to propose that our conclusions about the nature of the pairing in the superconductor and the nature of the normal state at $\nu=-2$ hold in TBG as well, with minor but important modifications. This allows for experiments in TTG, which in some respects are easier to interpret than those in TBG, to improve our understanding about both systems.
	
	What we do not attempt to do in this paper is a full-blown microscopic theory of TTG superconductivity. The questions we are interested in involve very complicated strongly interacting systems where the understanding from theory is only partially complete. Thus  we view our semi-phenomenological approach as the safest way of proceeding, since it allows us to avoid committing to any particular calculational schemes of specific  microscopic models, both of which come with uncertainties at present.

	\section{Basic model \label{sec:model}}
	
	Let us first review the basic theoretical description of TTG \cite{khalaf2019magic}. 
	We will take the graphene layers to lie normal to the $\uvz$ axis, and will gauge-fix the in-plane vector potential so that $\bfA_l = (2-l) \bfA$ on layer $l=1,2,3$, where $\bfA = \d\bfB\times \uvz$ with $\d$ the interlayer separation. 
	The displacement field $D$ enters the Hamiltonian through an onsite potential of $+U/2,0,-U/2$ on layers 1, 2, and 3 respectively, where $U \propto 2\d D/\ep_0$. The exact proportionality constant depends on details of how the screening works, and will not be important for the present discussion.
	
	To proceed it is most convenient to work in a basis of fields that have definite eigenvalues under the action of mirror $M$ \cite{khalaf2019magic,cualuguaru2021tstg}, which exchanges layers 1 and 3 and which is a symmetry when $U=\bfB = 0$. In the basis $(\psi_+,\psi_2,\psi_-)^T$ with $\psi_\pm = \frac1{\sqrt2}(\psi_1\pm \psi_3)$, where $\psi_l$ annihilate electrons on layer $l$,  the single-particle Hamiltonian is 
	\be \label{h0} H_0 = \bpm -iv_D\bfsig_v \cdot \D & \sqrt2 T &\cpl \\ \sqrt2T^\da & -iv_D\bfsig_v\cdot\D & 0 \\ 
	\cpl^\da & 0 & -iv_D\bfsig_v\cdot \D\epm  + H_Z,\ee 
	where the Zeeman energy $H_Z =  - \frac{\mu_Bg}2\bfB \cdot\bfs$.
	Here $v_D$ is the monolayer Dirac velocity, $T$ is the hopping matrix familiar from the bilayer problem (see appendix \ref{app:single_particle}), $ \cpl \equiv U/2-v_D \bfsig_v\cdot \bfA$, and we  have used the 
	notation $\bfsig_v \equiv (\tau^z\s^x,\s^y)$, with $\tau^a,\s^a,s^a$ Pauli matrices for valley, sublattice, and physical spin, respectively. 
	For simplicity we have also ignored a rotation of the $\bfsig_v$ on layers 1 and 3 by the twist angle, since its effects are small \cite{bistritzer2011moire,tarnopolsky2019origin}. Typical orbital magnetic energies are on the order of a few meV, with Zeeman energies smaller by a factor of $\sim3$.
	
	We note in passing that the Hamiltonian \eqref{h0} is the correct starting point only when the top and bottom graphene layers are not displaced relative to one another in the xy plane. A nonzero relative displacement $\bfr$ breaks mirror symmetry, and can have strong effects on the flat band physics \cite{lei2020mirror}. However, a vanishing relative displacement is likely to be favored energetically \cite{carr2020ultraheavy}, and at $\bfB=0$ the data of Ref. \cite{inplane_fields} appear to be mirror-symmetric to a good approximation. We will therefore set $\bfr = 0$ in the following. 
	
	From \eqref{h0}, we see that in the absence of $U$ or $\bfB$, $H_0$ decouples into a TBG Hamiltonian for the mirror-even fields $\psi_+,\psi_2$, which form a set of flat bands near the magic angle, together with a decoupled Dirac cone formed by the mirror-odd field $\psi_-$ \cite{khalaf2019magic}.
	Given that $U$ plays a crucial role in the observed superconductivity, the physics of the superconductivity is likely to be closely related to effects caused by the mirror-odd $\psi_-$ Dirac cone. 
	
	The mirror symmetry of the $U=\bfB=0$ Hamiltonian has other important consequences for how $\bfB$ affects the single-particle physics, as both $U$ and $\bfB_{\rm orb}$ are odd under mirror (here $\bfB_{\rm orb}$ denotes the orbital field). In particular, at $U=0$ the spectrum at each momentum must be an even function of $B_{\rm orb}$.
	This means that unlike in TBG, at $U=0$ there will be no field-induced de-pairing of intervalley Cooper pairs due to misalignement of the Fermi surfaces in the two valleys \cite{cao2020nematicity}, as here the single-particle energy satisfies $\ep_K(\bfk) = \ep_{K'}(-\bfk)$ even when $B_{\rm orb} \neq 0$. The fact that the leading effects are at order $B_{\rm orb}^2$ also means that at small $U,B$, the orbital effects of the magnetic field are rather weak.
	Diagonalizing $H_0$ at realistic values of $U,\bfB$, one finds two Dirac cones at the moire $K$ point at energies separated by an amount which increases with $U,B$, as well as a Dirac cone located very near the $K'$ point. Details on the single-particle physics can be found in appendix \ref{app:single_particle}. 
	
	One observation from Ref. \cite{inplane_fields} is that despite the sensitivity of the observed superconductivity to $U$ and $\bfB$, superconductivity near optimal doping is seen even when $U=\bfB=0$. Therefore the mirror-even fermions $\psi_+,\psi_2$ must be superconducting on their own, independent of their coupling to the mirror-odd $\psi_-$.\footnote{One possibility is that dissipation from the $\psi_-$ fermions promotes superconductivity in the mirror-even sector, along the lines of the mechanism in Ref. \cite{vishwanath2004screening}. In what follows we will however assume that the dominant effects of the $\psi_-$ fermions arise from the terms written in \eqref{masterl}.}
	We are therefore prompted to analyze the system by considering the Lagrangian (suppressing all indices)
	\bea \label{masterl}\mcl& =\mcl_{SC}[\psi_+,\psi_2,\De] + \mcl_-[\psi_-,\De] + \mcl_\pm[\psi_+,\psi_-],\\
	\mcl_- & =  \psi_-^\da (\p_\tau + v_D \bfk \cdot \bfsig_v  )\psi_- + \l (\psi_-\De^\da \psi_- + \psi_-^\da \De \psi_-^\da) \\ 
	&\qq - \frac{\mu_Bg}2  \psi_-^\da \bfB \cdot\bfs\psi_- \\ 
	\mcl_\pm & =  \psi_+^\da \cpl \psi_-  + \psi_-^\da \cpl^\da \psi_+,\eea 
	where $\De$ is the superconducting order parameter and $\mcl_{SC}[\psi_+,\psi_2,\De]$ is some unknown Lagrangian capturing the physics of the $U=\bfB=0$ superconductor, together with the Zeeman energy for the mirror-even fermions. The term proportional to $\l$ arises from a mean-field decoupling of an interaction of the form $\psi_{+/2}^\da \psi_{+/2}^\da \psi_-\psi_- + h.c$. Such a term does not arise from a standard density-density interaction, but as it is allowed by symmetries we will keep it. 
	
	Importantly, the modifications to the superconductivity by $U, B_{\rm orb}$ can be reliably addressed by integrating out the $\psi_-$ fermions perturbatively, as it is only through the mixing with these fermions that these perturbations have any effect. 
	
	\section{The normal state near $\nu=-2$ \label{sec:normal_state}}
	
	Before attempting to obtain an effective action for $\De$ by integrating out the fermions in \eqref{masterl}, it is helpful to think about constraints on the matrix structure of $\De$ by considering the likely nature of the $D=0,B=0$ normal state near $\nu = -2$. As we will see, known properties of the superconductor will help us place nontrivial constraints on the nature of the normal state. Because this discussion takes place at $D=0,B=0$, we may temporarily ignore the mirror-odd $\psi_-$ fermions: this is because in this limit the leading coupling of the $\psi_-$ fermions to the mirror-even $\psi_+,\psi_2$ fields vanishes,\footnote{In principle there are also higher-body interactions between the $\psi_-$ fermions and the mirror even sector --- in what follows we will assume that these interactions do not qualitatively change the nature of the normal state, which seems reasonable on account of the low density of the $\psi_-$ fermions.} rendering the problem equivalent to that of TBG \cite{khalaf2019magic}. This simplication is quite useful, as it allows us to draw on insights gained in the study of TBG. Aspects of the physics particular to TTG will arise when we turn on nonzero $D,B$, which is discussed in the following sections.
	
	Existing experiments on TTG \cite{park2021tunable,hao2021electric} indicate that the normal state at $\nu=-2$ is likely highly resistive (and presumably would be insulating in the absence of the $\psi_-$ Dirac cone). Additionally, the doped state at $\nu = -2-\delta$ is seen to possess a Landau level  degeneracy reduced from that of the charge neutrality point by a factor of two.   
	The simplest way to produce a ground state in accordance with these observations is to polarize a flavor, and then to open an interaction-induced gap in the remaining polarized subspace. 
	We will be making the assumption that the superconductor pairs fields related by time reversal, implying that both $K,K'$ valleys participate in the superconducting state. We will also be assuming (at least at small $B,T$) that the superconductor inherits the flavor polarization of the $\nu = -2-\d$ normal state. This implies that the normal state is not valley polarized, and in what follows we will assume that the polarized flavor is spin. Spin polarization here means that the ground state is spin-valley locked (SVL), with the spins in a given valley fixed to lie along some (valley-dependent) direction.  
	
	The assumption of spin-valley locking is reasonable from a theoretical point of view, as polarizing spin (as opposed to valley) allows the system to take advantage of the intervalley Hund's coupling. This assumption is also motivated by experiment, as SVL allows the superconductor to achieve the strong Pauli limit violation observed in Ref. \cite{inplane_fields}.
	
	The simplest way to open up a gap is to give some fermion bilinear $c^\da_\a Q_{\a\b} c_\b$ a nonzero expectation value, where the indices $\a,\b$ run over the states in the low-energy SVL subspace. 
	To determine what we should take for the form of $Q$, we will borrow results from the strong-coupling picture proposed for twisted bilayer graphene in Refs. \cite{bultinck2020ground,khalaf2020charged}. In this picture, the four flat bands in the SVL subspace can be labeled as  $|K,A,\xi\ran,|K',B,\eta\ran$ and $|K',A,\eta\ran, |K,B,\xi\ran$, where $\k\xi,\k\eta$ denote normalized spinors determining the directions along which the spins are locked in the $K$, $K'$ valleys respectively, and where $A/B$ is the sublattice index (which is a good index for the flat bands in the strong-coupling limit \cite{bultinck2020ground}). The first set of two bands has Chern number $C=\s^z \tau^z = +1$, and the second set of two have $C=-1$.
	The strong-coupling analysis indicates that $Q$ must mix the two valleys, and must be proportional to $\s^y$ in sublattice space, this being the only matrix structure which allows the ordered state at $\nu=-2$ to completely fill two of the Chern bands and minimize the kinetic and potential energy in the strong-coupling limit (see \cite{bultinck2020ground} for the full logic). Any $Q$ which fulfills these properties and survives projection onto the occupied SVL subspace may be written as 
	\be Q =  \s^y \bpm & |\eta\ran\lan \xi | e^{i\vp} \\ 
	|\xi\ran\lan \eta| e^{-i\vp} & \epm,\ee 
	where the matrix written out explicitly is in valley space and where the phase $\vp$ is arbitrary. A nonzero expectation value for $c^\da_\a Q_{\a\b}c_\b$ breaks time reversal symmetry $\mct = s^y \tau^x \mck$ unless the spins in each valley are antiferromagnetically aligned.

	At zero magnetic field, the relative orientation of $\k\xi,\k\eta$ is dictated by the sign of the intervalley Hund's coupling $J_H$. The sign of $J_H$ is not known a priori, and is determined by a delicate competition between various effects, e.g. Coulomb interactions and coupling to phonons \cite{chatterjee2020symmetry}. However, as we will see momentarily, the fact that the superconductor seen in experiment has a nonzero BKT transition temperature at zero field means that $J_H$ must favor antiferromagnetic alignment of the spins in the two valleys (for the case without SVL, see appendix \ref{app:lg_details}). For definiteness we will take the zero-field normal state to have $\k\xi = \k\upa, \k\eta=\k\doa$, so that 
	\be Q = \s^y (\tau^+ s^+e^{i\vp} + \tau^-s^- e^{-i\vp}).\ee 
	
	There are two important things to note about $Q$. The first is that it is invariant under time reversal symmetry, as the spins are aligned antiferromagnetically.
	Given that the zero-field state at $\nu=-2$ contains two filled bands with Chern numbers $\pm1$, this antiferromagnetic SVL state  is, remarkably,  therefore a topological insulator protected by the unbroken $U(1)$ and $\mct$ symmetries.\footnote{This should be contrasted with the analysis of the spinless model of Ref. \onlinecite{bultinck2020ground} where the proposed $K-IVC$ state is also topological, protected by $U(1)$ and a $\mct'$ symmetry  which combines  time reversal with  a valley rotation. Such a symmetry is not sufficient to protect gapless edge states as the valley symmetry is broken at the edge. In contrast the spinful state discussed in this paper will have protected edge states. } 
	The second is that $SO(2)_s$ rotations of the spin about $\uvz$ act on $Q$ in the same way as the group $U(1)_o$ of opposite phase rotations in the two valleys. Therefore the order parameter manifold of the insulator is $(U(1)_o \times SO(3)_s)/U(1)_{o+s} = SO(3)$. As $\pi_2(SO(3)) = 0$ there are no skyrmions in this state (only $\zt$ vortices), complicating a skyrmion-driven mechanism for superconductivity along the lines of that proposed in Refs \cite{khalaf2020charged,chatterjee2020symmetry} (although depending on the strength of $J_H$, a description in terms of skyrmions that exist at intermediate scales may still be useful).

	\section{Ginzburg-Landau theory \label{sec:lg}} 
	
	We now turn to discussing the general features that we expect any Ginzburg-Landau (GL) free energy for the observed superconductor should satisfy.

	\begin{figure}
		\includegraphics{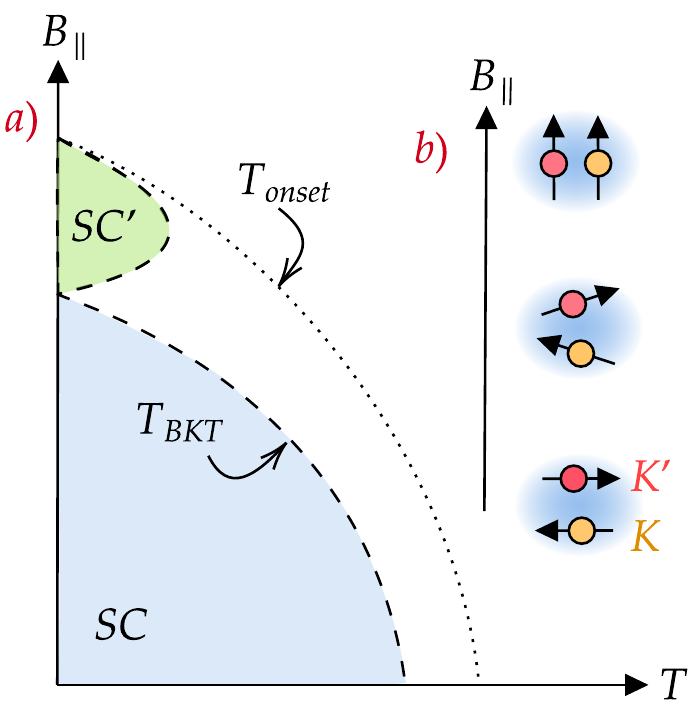}
		\caption{\label{fig:inspir_schematic} a) A schematic of the phase diagram in the $B_\prl$-$T$ plane. SC denotes the low-field superconductor, and SC' denotes the re-entrant finite-momentum pairing phase. The inset b) shows a schematic of how the pairing evolves in the field: the electrons in the Cooper pair have their spins oppositely aligned in the $K,K'$ valleys in zero field, with the spins smoothly rotating to point along the field at large $B_\prl$.  } 
	\end{figure}
	
	As mentioned above, we are operating under the assumption that $\De$ is off-diagonal in the valley index. 
	Interaction effects similar to those mentioned above then mandate \cite{khalaf2020charged,khalaf2020symmetry,lee2019theory} that $\De$ be proportional to the identity in sublattice space (so that pairing takes place between opposite Chern sectors). For the present discussion, we will further assume that over the full temperature range of interest, the spin-valley locking which likely occurs in the normal state is inherited by the superconductor (the case where this assumption is relaxed is treated in appendix \ref{app:lg_details}). 
	
	With these assumptions, we may write the order parameter as 
	\bea \De &= \frac{|\De|}2  \bpm & -|\eta\ran \lan \xi^*|\\ 
	|\xi\ran \lan \eta^* |& \epm \\ 
	& = \frac{|\De|}{2^{3/2}}\(i\tau^y \bfd \cdot \bfs  + \tau^x d_0\) is^y,
	\eea 
	where the matrix on the first line is in valley space, and the $d$ four-vector has components 
	\be d^\mu = \frac1{\sqrt2}\lan  \xi^*| is^y s^\mu |\eta\ran.\ee 
	Note that in writing down this form for the order parameter, we have assumed that the pairing occurs in an even angular momentum channel. None of the analysis to follow is affected by this choice, and odd angular momentum pairing can be treated simply by interchanging $\tau^x$ and $\tau^y$ in the above expression for $\De$.
	
	A general expression for the GL free energy in the SVL subspace can be written in terms of $\De$ and the vectors $\bfxi = \lan \xi | \bfs |\xi\ran, \, \bfeta = \lan \eta | \bfs | \eta\ran$ as 
	\bea \label{free_energy} f & = \frac{c}2 |\p \De|^2 + r|\De|^2 + \frac{u}2|\De|^4 +  w |\De|^2\bfeta \cdot \bfxi  \\ & + \kappa|\De|^2 (\bfB\cdot(\bfxi-\bfeta))^2- \zeta  |\De|^2 \bfB \cdot(\bfxi+\bfeta),\eea 
	where $-r,u,\kappa>0$, the coefficient $w$ determines whether the spins in each valley prefer to align ferromagnetically or antiferromagnetically, 
	and where all of the coefficients in the above expression contain implicit dependence on $U$ and $B$, with this dependence determined from the Lagrangian \eqref{masterl} (to be discussed shortly).
	
	Suppose first that $w<0$, so that the spins in each valley prefer to align ferromagnetically along some direction $\bfn$, which is chosen spontaneously when $\bfB=0$. The order parameter is then invariant under the combined action of $U(1)_g$ gauge rotations and $SO(2)_s$ spin rotations about $\bfn$. The order parameter manifold for the zero-field superconductor is therefore $(U(1)_g \times SO(3)_s)/U(1)_{o+s} = SO(3)$. Consequently, this phase does not have a finite-$T$ BKT transition at $\bfB=0$. Furthermore, since this phase only has $\zt$ vortices, the critical current in zero field vanishes \cite{cornfeld2020spin}, and when $\z\neq 0$ the free energy of this state decreases linearly in the presence of a small in-plane field. All of these features are in disagreement with experiment.
	
	When $w>0$ on the other hand, the spins align antiferromagnetically in the plane normal to 
	the field direction at small $B$, and then gradually cant to point along the field at larger $B$ (see Fig. \ref{fig:inspir_schematic}). This canting process happens without breaking apart the Cooper pair, allowing the superconductor to survive at fields above the Pauli limit.
	Unlike in the $w<0$ case, the free energy and condensate fraction do not evolve linearly with $B$ near $B=0$ (appendix \ref{app:lg_details} contains the details). 
	Furthermore, since $\De$ is invariant under $SO(2)_s$ spin rotations about the ordering direction when $\bfB=0$, the order parameter manfiold in the zero field limit is $(U(1)_g \times SO(3)_s) / (\zt\times SO(2)_s) = U(1)\times S^2/\zt$. Then there are vortices with integer-valued charge that cost logarithmic energy, leading to a nonzero $T_{BKT}$ at zero field. All of these features are in agreement with experiment.
	
	We are therefore forced to conclude that the zero-field superconductor favors antiferromagnetic alignment. This provides the justification for our assumption in the previous section that the normal state has antiferromagnetic spin-valley locking (see Ref. \cite{khalaf2020symmetry} for a similar argument). 
	
	It now remains to understand the mechanism producing the re-entrant phenomenon seen in experiment, as well as to understand the dependence of $T_{\rm onset}$ and $T_{BKT}$ on the applied fields. This can be done by understanding the $U,B$ dependence of the coefficients appearing in the free energy, which we now turn to discussing. 
	
	\section{Quantum Lifshitz criticality and finite-momentum pairing \label{sec:lifshitz}}
	
	As mentioned in the introduction, a particularly noteworthy aspect of the experiment is that $T_{\rm onset}$ decreases more or less smoothly all the way until superconductivity is completely killed, while $T_{BKT}$ shows non-monotonic behavior in $B_\prl$ \cite{inplane_fields}.\footnote{We note however that a subsequent experiment \cite{liu2021coulomb} did not find evidence for a non-monotonic $T_{BKT}$.} The temperature 
	$\ton$ is set by the $T = 0$ condensate fraction $\lan |\De|^2\ran$. By contrast, $\tb$ is set by the $T = 0$ superfluid stiffness $K$, viz. the coefficient of the $(\p \phi)^2$ term in the free energy, where $\phi$ is the phase mode of $\De$. 
	
	A natural explanation for the observed re-entrant phenomenon is therefore that $\r$ decreases monotonically with $B$ by a factor $\propto \sqrt{1-B^2}$, while $K$ decreases {\it faster} with $B$, and goes to zero when $B=B_{\rm re-entrant}$, where the phase transition occurs (see the schematic phase diagram in Fig. \ref{fig:inspir_schematic}.) When $B>B_{\rm re-entrant}$, $K$ becomes negative. 
	
	The IR physics of the superconductor is dominated by the phase mode $\phi$,\footnote{For nonzero field strengths which are not strong enough to fully lock the spins along $\bfB$, there is an additional neutral spin mode coming from $U(1)$ spin rotations about the field direction. It however is not important for the following discussion, and will be ignored.} with a Lagrangian capturing the essential physics being 
	\be \label{lifshitzl} \mcl = \frac12 \((\p_\tau\phi)^2 + K (\p\phi)^2 + \vs (\D^2\phi)^2\),\ee 
	where $\vs>0$ is included for stability. The proposed re-entrant transition is a Lifshitz point, where $K$ vanishes. 
	
	Whether or not a Lifshitz transition can occur is determined by the field dependence of the coefficients in the GL free energy, which can be calculated by integrating out the $\psi_-$ fermions in the Lagrangian \eqref{masterl}. The important point here is that $U,B$ only enter \eqref{masterl} in the coupling of $\psi_+$ to $\psi_-$, and since $\psi_-$ is simply a free Dirac cone, the field dependence of the terms in the free energy can be calculated in a perturbatively controlled way. 
	We find that to quadratic order in $U$ and $B$, the coupling between the $\psi_-,\psi_+$ fermions result in the parameter $r$ and the stiffness $K$ appearing in the free energy \eqref{free_energy} being renormalized by amounts 
	\bea\label{drdk} \d r & = -aU^2 + bB^2 \\ 
	\d K & = cU^2 - dB^2,\eea 
	where $a,b,c,d$ are all positive coefficients, whose detailed form is not important for the present discussion. The upshot is that (at least at small fields), the displacement field tends to promote superconductivity, while the orbital field leads to a suppression in both the condensate fraction and spin stiffness. 
	The fact that the orbital and displacement fields have opposite effects on the superconductor (at least at small $U,B$) originates from the different ways in which they couple the $\psi_+,\psi_-$ fermions together. Importantly, this is exactly the behavior seen in experiment, with the superconductivity strengthened (suppressed) with increasing $U$ (increasing $B$), at least in the range of displacement fields for which the re-entrant transition occurs.	
	The details of the calculation leading to \eqref{drdk} are standard, and are relegated to appendix \ref{app:lg_coeffs}.
	A rough estimate of the critical in-plane field for the superconductor at $U=0$ in terms of the mircoscopic parameters which enter the coefficients appearing in \eqref{drdk} is given in appendix \ref{app:best}. Our estimate is of the same order of magnitude as the critical field observed in experiment and we will be satisfied with this consistency check, leaving a more detailed calculation to future work.
	
	This dependence of $\d r,\d K$ on $B^2$ is precisely of the form required for the proposed Lifshitz transition to work, with the Lifshitz transition occurring if $d$ is large enough to drive $K$ to zero before $r$. Furthermore, the fact that the condensate fraction increases with $U$ to quadratic order provides an explanation for why, at moderate values of $U$, the superconductivity is seen to be stronger with increasing $D$. That the re-entrant transition occurs only at moderate values of $D$ can be understood in the same way, as the Lifshitz transition can only occur if $\langle |\Delta|^2\rangle$ remains nonzero at the field strength where $K$ is driven to zero. In this scenario, at small values of $D$ it is $\langle|\Delta|^2\rangle$ which goes to zero first with increasing $B$, while at larger values of $D$ it is $K$ that goes to zero first. 
	
	Note that due to the rotational symmetry breaking by the in-plane field, it is possible that the stiffnesses along and normal to the field direction differ, so that 
	\be \mcl = \frac12 (K_\prl (\p_\prl \phi)^2 + K_\perp (\p_\perp \phi)^2) + \cdots,\ee 
	where $\p_\prl$ ($\p_\perp$) denote derivatives along (normal to) the field direction. One could therefore consider a situation in which only one of $K_\prl,K_\perp$ is driven to zero at the Lifshitz transition. However, the isotropic case where $K_\prl,K_\perp$ go to zero simultaneously seems more likely to be realized. Indeed, consider the case where only one component vanishes at $B_{\rm re-entrant}$. The critical point is an anisotropic Lifshitz model. In appendix \ref{app:lifshitz} we show that this model orders at $T=0$, and displays QLRO up to some non-universal nonzero temperature. Thus in this scenario we would {\it not} expect the observed $T_{BKT}$ to go all the way to zero at $B_{\rm re-entrant}$ (although the temperature at which the anisotropic model looses QLRO is likely very small, and may well be below the current experimental resolution). 
	
	On the other hand consider the isotropic case, where both $K_\prl$ and $K_\perp$ vanish at $\bre$. In this case the critical point has no QLRO at any temperature \cite{ghaemi2005finite}, so that $T_{BKT}=0$ at the critical point, as seems to be the case experimentally. Another reason for favoring the isotropic case is that to leading order in $B$, $K_\prl$ and $K_\perp$ are renormalized by the same amount.
	Our prior is therefore that near the critical point, $K_\perp = K_\prl$ to a good approximation. However, it may well be the case that in narrow field regimes only one of $K_\perp,K_\prl$ is negative, possibly explaining the rather broad nature of the transition and the observed small pockets of potentially distinct superconducting phases \cite{inplane_fields}. 
	
	On the other side of the Lifshitz phase transition, the pairing occurs at finite momentum. Note that this scenario is very different in spirit to the usual FFLO mechanism \cite{ff,lo}, since the latter relies on Pauli pair breaking being the dominant mechanism to suppress superconductivity. This is opposite to case of the superconductor currently under consideration, which strongly violates the Pauli limit and in which the main effects of the field are orbital in nature. Because the present mechanism differs from that of FFLO, in order to minimize the condensation energy the order parameter will presumably live at a single nonzero wavevector $\mathbf{q}\neq0$, so that that $\Delta(\bfr) \sim e^{i \bfq\cdot\bfr}$ has a spatially uniform magnitude.
	
	The first indications from Ref. \cite{inplane_fields} are that the re-entrant transition is first order. 
	As written down in \eqref{lifshitzl} the transition is second order, but in general it may be driven first order by other higher-order terms not included in \eqref{lifshitzl} \cite{fradkin2004bipartite,vishwanath2004quantum}. One example is the term  $\l (\D\phi\cdot \D\phi)^2$, which is marginally relevant if $\l<0$ \cite{vishwanath2004quantum}.\footnote{The cubic term $\p_x \phi (\p_x \phi - \sqrt3\p_y\phi)(\p_x\phi + \sqrt3\p_y\phi)$ also renders the critical point first order, but is not invariant under the $C_6$ symmetry of the theory, which sends $\psi(\bfx) \mt \tau^x\s^x e^{2\pi i \tau^z\s^z}\psi(R_{\pi/3}\bfx)$ and $\bfB \mt R_{\pi/3}\bfB$ \cite{po2018origin}. Also note that nonzero $\bfB$ permits us to write down other relevant cubic terms like $UB^i\p_i\phi (\p\phi)^2$ (with the $U$ included to preserve mirror symmetry). These terms however are not generated in the context of our model, as the effects of the orbital coupling only enter at order $B^2$ (see appendix \ref{app:lg_coeffs}).}

	\section{Lessons for TBG}

	As was emphasized above, in the absence of external fields the $\psi_-$ fermions decouple form the mirror-even sector, rendering the system nearly identical to TBG together with a decoupled Dirac fermion. In particular, this means that the  physics  of the superconductor and the normal state near $\nu=-2$   are likely to be the same in both TBG and TTG.   As we have argued that at $D=0,B=0$ the superconductor near $\nu=-2$ in TTG has antiferromagnetic spin-valley locking and pairing which involves a superposition of spin singlet and spin triplet, the same should be true in TBG (see also Ref. \onlinecite{khalaf2020symmetry}) . 
		
		How should we reconcile this conclusion with the striking differences between TBG and TTG   in the presence of a nonzero in-plane magnetic field?  In TBG, the critical in-plane field in the superconducting state at optimal filling is $B_{\prl,c}\approx 1T$ \cite{cao2018unconventional}, which is much smaller than in TTG, and is much closer to the Pauli limit.  Since we claim that the pairing in TBG also involves a combination of spin singlet and spin triplet, the comparatively small value of $B_{\prl,c}$ in TBG cannot be due to Zeeman effects. Instead, the critical field is set by orbital effects. Indeed, unlike in TTG the single-particle energies in TBG do {\it not} satisfy $\ep_K(\bfk) = \ep_{K'}(-\bfk)$ in a nonzero orbital field, allowing orbital pair-breaking effects to produce a relatively lower value for $B_{\prl,c}$ \cite{cao2020nematicity}.  The suppression of superconductivity at roughly the Pauli limit in an in-plane field was interpreted early on as evidence for a spin singlet superconductor. This interpretation was first called into question by the observation that the in-plane critical field is strongly angle dependent \cite{cao2020nematicity}. Such angle dependence cannot arise from Zeeman effects. Within the present picture the superconductor is a superposition of singlet and triplet, its  suppression by an in-plane field is entirely orbital, and the rough agreement with the Pauli limit is a coincidence coming from the rough equality of the orbital Zeeman  g-factor and the spin g-factor.
		
		Turning to the normal state of TBG, a striking consequence of the antiferromagneic spin-valley locking is that, within the strong coupling theory, the correlated insulating state at $\nu = -2$ is an interaction-driven $\zt$ TI. This state will have time reversal protected edge states which will be interesting to look for in future experiments. 
		
		Finally, a puzzle which we do not resolve is that the {\it insulating} state at $\nu=-2$ in TBG is also suppressed by an in-plane magnetic field \cite{cao2018correlated,yankowitz2019tuning} (while in TTG the suppression seems to be mostly absent \cite{liu2021coulomb}). There are conflicting reports \cite{yankowitz2019tuning,wu2021chern,park2021flavour} on the  details of how the insulating gap in TBG evolves with a field  (linear in some experiments, quadratic in others), and we leave the understanding of this phenomenon to the future.

	\section{Discussion and Summary}

	In this paper we have used a combination of  phenomenological and microscopic theoretical reasoning to build a coherent picture of the superconductivity and related phenomena in TTG. We argued that the superconducting state is a superposition of spin singlet and spin triplet pairings. We showed how this explains the contrasting effects of small displacement and $B_\parallel$ fields, even though both break mirror symmetry. We also showed that the mirror symmetry explains the striking difference between the stability of TTG and TBG superconductors to in-plane fields despite their close microscopic relationship. We proposed that the re-entrant superconductivity recently observed in twisted trilayer graphene \cite{inplane_fields} can be explained by a spin-valley locked superconductor which undergoes a quantum Lifshitz transition into a finite-momentum pairing state in the presence of an in-plane magnetic field. The close relationship between TBG and TTG when the displacement field and magnetic field are turned off enables insights from the bilayer system to contribute to the study of the trilayer case, and vice versa. One consequence of this is our conclusion that the normal state near $\nu=-2$ in twisted bilayer graphene may be a $\mathbb{Z}_2$ TI.
	
	A possible test of our proposal for the re-entrant superconductor in TTG is to measure the Josephson current between two neighboring TTG regions maintained at different displacement fields. Due to the dependence of $B_{\rm re-entrant}$ on the displacement field $D$, it is possible to tune between the low-field and high-field phases in constant magnetic field, by varying $D$ alone. Therefore one could imagine engineering a Josephson tunneling experiment between the low- and high-field superconducting states simply by varying the applied electric field in space. If the high-field state has finite-momentum pairing, the Josephson current should vanish, due to the spatially oscillating phase of the high-field phase's order parameter \cite{yang2000josephson}. Of course, a vanishing current could also occur by virtue of the high- and low-field phases having different pairing symmetries. However, if the pair momentum in the high-field state is a continuously varying  function of $U$ (as in our analysis), the Josephson current between two high-field superconducting regions maintained at different displacement fields should also vanish, providing a stronger test of the proposed Lifshitz scenario.
	
	In the case which the high-field state is obtained through an (anisotropic) Lifshitz transition, one could also simply measure whether or not the critical current is direction-dependent, as near the phase transition the critical current should be highly direction-dependent (see appendix \ref{app:lifshitz}). 
	
	Finally, the conclusion that the normal state near $\nu = -2$ is a $\zt$ TI is also interesting from an experimental perspective, as it will have symmetry protected gapless edge states in a single domain sample. In TTG, this TI coexists with the massless Dirac fermions of the mirror-odd sector.  Complications caused by the mirror-odd Dirac cone could potentially be avoided in a setup where the sample edge is mirror symmetric and electrical contact is made only with the middle layer, enabling the transport properties of the mirror-even sector to be isolated. It may also be interesting to study the effects of induced spin orbit coupling in this system, along the lines of Ref. \cite{lin2021proximity}, to explicitly lock in the symmetry breaking leading to the TI sector.   
	
	The proposed $\zt$ TI at $\nu = -2$ may be more readily accessible to experiments in TBG, where there are no extra gapless mirror-odd fermions.

	\section*{Acknowledgements}
	
	We thank Yuan Cao, Jane Park, and Pablo Jarillo-Herrero for extensive conversations about their experiments. EL is supported by the Fannie and John Hertz Foundation and the NDSEG fellowship. TS was supported by US Department of Energy grant DE- SC0008739, and partially through a Simons Investigator Award from the Simons Foundation. This work was also partly supported by the Simons Collaboration on Ultra-Quantum Matter, which is a grant from the Simons Foundation (651440, TS).  
	
	\newpage 
	
	\begin{widetext}
		
		\appendix

		\section{Microscopic derivation of the terms in the Ginzburg-Landau free energy \label{app:lg_coeffs}}

		In this appendix we describe how the $\bfB$ and $U$ dependence of the coefficients appearing in GL free energy \eqref{app_full_lg} can be obtained microscopically, starting from the Lagrangian 
		\bea\mcl& =\mcl_{SC}[\psi_+,\psi_2,\De] + \mcl_-[\psi_-,\De] + \mcl_\pm[\psi_+,\psi_-],\\
		\mcl_-[\psi_-,\De] & =  \psi_-^\da (\p_\tau + v_D \bfk \cdot \bfsig_v  )\psi_- + \l (\psi_-\De^\da \psi_- + \psi_-^\da \De \psi_-^\da)  - \frac{\mu_Bg}2  \psi_-^\da \bfB \cdot\bfs \psi_- \\ 
		\mcl_\pm[\psi_+,\psi_-] & =  \psi_+^\da \cpl \psi_-  + \psi_-^\da \cpl^\da \psi_+,\eea 
		where as in the main text $\cpl = U/2 - v_D \bfA \cdot \bfsig_v$ (recall that $\bfsig_v = (\tau^z\s^x,\s^y)$), and where $\mcl_{SC}[\psi_+,\psi_2,\De]$ contains the physics of the mirror-even flat band fermions and the $U=B=0$ superconductor. As in the main text, we will assume even angular momentum pairing and write the order parameter as 
		\bea\label{deltadecomp} \De & =  i\tau^y\, \bfd \cdot \bfs\, is^y  + \tau^x \, d_0\, is^y.\eea  
		The fact that $\De$ is proportional to the identity in the sublattice index and off diagonal in the valley index is important, with this matrix structure being responsible for several important minus signs in what follows. However, the assumption of even angular momentum pairing is not important, and odd angular momentum pairing can be treated simply by exchanging $\tau^x$ and $\tau^y$ in the above expression.
		
		In what follows we will treat the mirror-even fermions $\psi_+,\psi_2$ by assuming that they form a flavor unpolarized Fermi liquid. Situations in which flavor polarization occurs can be easily dealt with by inserting appropriate projectors into traces. In particular, if the polarized flavor is spin, with the spins being locked to valley (which as we have argued in the main text is likely the case relevant for experiment), the effects of the polarization only enter when dealing with the Zeeman couplings, and can be dealt with in the manner described near \eqref{proj_props}.

		\ss{Orbital effects} 
		
		We begin by examining the orbital effects of the magnetic field. The only place where the orbital coupling enters into the above Lagrangian is in the $\mcl_\pm$ term, which couples the symmetric and antisymmetric sectors. This allows the effects of $U,B$ to be treated within a controlled perturbative expansion.
		
		Integrating out $\psi_+$ produces the term 
		\be \d \mcl =- \frac12 \bpm \psi_-^\da & \psi_-\epm \bpm \cpl^\da & \\ & -\cpl^T \epm \mcg_+ \bpm \cpl & \\ & -\cpl^*\epm \bpm \psi_- \\ \psi_-^\da \epm,\ee 
		where $\mcg_+$ is the $\psi_+$ propagator. Since  we are remaining largely agnostic about the physics of the symmetric flat bands, we do not know what $\mcg_+$ looks like, other than that it involves $\De$s on the off-diagonals. In what follows we will simply assume that the most important effects of $\d\mcl$ can be captured by the term 
		\bea \label{hdef} \d \mcl & \supset  h \( \psi_-^\da \cpl^\da \De \cpl^* \psi_-^\da + \psi_- \cpl^T \De^\da \cpl\psi_- \) \eea 
		where $h$ is a constant. Here and below, the notation $A\supset B$ is to be read as ``$B$ is a term appearing in $A$''. 
		
		We now integrate out $\psi_-$ and examine the structure of the resulting terms generated in the effective action $S[\De]$ for $\De$. Thus the remaining task is to compute the functional determinant 
		\bea S[\De] & \supset 
		-\Tr \ln \[ \unit +  \bpm \frac1{-i\o + \ep} & \\& \frac1{-i\o - \ep^T} \epm \bpm & \sch \\ \sch^\da & \epm \]
		\\ 
		&=  \sum_{n>0} \frac{1}{2n} \Tr\[ \( \bpm  G_p& \\& G_h \epm \bpm & \sch \\ \sch^\da & \epm\)^{2n} \] \\ 
		& =  \sum_{n>0} \frac1{n}\Tr\[( G_p \sch G_h \sch^\da)^{2n} \],   \eea
		where 
		\be \sch = h\cpl^\da \De \cpl^* + \l \De \ee 
		and where the particle / hole Greens functions are 
		\be G_p = \frac1{-i\o + \ep},\qquad G_h = \frac1{-i\o - \ep^T}.\ee 
		We will assume that integrating out $\psi_+$ leaves the $\psi_-$ fermions as continuing to be well-described by a Dirac cone, albeit one with a renormalized (possibly anisotropic) Dirac velocity and a nonzero chemical potential. Therefore we take $\ep$ to be (ignoring for now the Zeeman coupling) 
		\be \ep = v_x k_x  \tau^z \s^x + v_y k_y  \s^y - \mu,\ee 
		so that 
		\be G_p = \frac{i\o +\mu + v_xk_x\tau^z\s^x + v_y k_y \s^y}{-(-i\o -\mu)^2 + (v_xk_x)^2+(v_yk_y)^2},\qq  G_p = \frac{i\o -\mu + v_xk_x\tau^z\s^x - v_y k_y \s^y}{-(-i\o +\mu)^2 + (v_xk_x)^2+(v_yk_y)^2}.\ee 
		
		In the following we will use the notation 
		\be \mcl[\De] = \sum_{n>0} \mcl_{2n}[\De],\ee 
		with $\mcl_{2n}$ containing $2n$ powers of $\De$. We will first turn to examining the quadratic piece $\mcl_2[\De]$. 
		
		\sss{Quadratic terms} 
		
		We will work at $\mu=0$ for now, as the effects of a small nonzero $\mu$ do not qualitatively modify the structure or field dependence of the terms that are generated. We will also only focus on the behavior of $\mcl_2[\De]$ in the zero-frequency limit, by taking $\De = \De(0,\bfq)$. After re-scaling momenta by $k_x \mt k_x /v_x$, $k_y\mt k_y/v_y$, we have
		\bea 
		\mcl_2[\De(0,&q_x/ v_x,q_y/v_y)] \supset +\frac{1}{v_xv_y} T \sum_n \int_{\bfk} \Tr \Big[\frac{i\o_n + \bfk\cdot\bfsig_v }{\o_n^2 + k^2} (h\cpl^\da \De \cpl^*+\l\De) \frac{i\o_n + (\bfk+\bfq) \cdot\bfsig_v^*}{\o_n^2 + (\bfk+\bfq)^2} (h\cpl^T \De^\da \cpl+\l\De^\da) \Big] \eea 
		
		To perform the trace, we can use the explicit expression for $\cpl$ to write 
		\bea
		\Tr[\bfk \cdot \bfsig_v \De (\bfk+\bfq)\cdot\bfsig_v^* \De^\da] & = - \bfk\cdot(\bfk+\bfq) |\De|^2 \\
		\Tr[ \cpl^\da \De \cpl^* \De^\da ] & =\G |\De|^2\\
		\Tr[\bfk\cdot \bfsig_v \cpl^\da \De \cpl^* (\bfk+\bfq)\cdot\bfsig^*_v\De^\da] 
		& = -\G (\bfk+\bfq)\cdot\bfk |\De|^2 \\ 
		\Tr[ \cpl^\da \De \cpl^* \cpl^T \De^\da  \cpl ] & = \G^2|\De|^2 \\
		\Tr[\bfk\cdot\bfsig_v\cpl^\da \De \cpl^*(\bfk+\bfq)\cdot\bfsig_v^* \cpl^T \De \cpl ] & = -\G^2 \bfk\cdot(\bfk+\bfq) |\De|^2,
		\eea 
		where in this appendix $|\De|^2 = \Tr[\De^\da \De]$, we have defined 
		\be \G \equiv \frac{U^2}4 - v_D^2 A^2,\ee 
		and have used $\bfsig_v \De = -\De \bfsig_v^*$ as well as 
		\bea \Tr[ \bfa \cdot\bfsig  \bfb \cdot\bfsig \bfc\cdot\bfsig\bfd\cdot\bfsig] & =( (\bfa\cdot\bfb)(\bfc\cdot\bfd) - (\bfa\times\bfb)\cdot(\bfc\times\bfd))\Tr[\unit]\\ 
		& = ((\bfa\cdot\bfb)(\bfc\cdot\bfd) -(\bfa\cdot\bfc)(\bfb\cdot\bfd) + (\bfa\cdot\bfd)(\bfb\cdot\bfc))\Tr[\unit].\eea 
		The fact that $A$ and $U$ enter in $\G$ with opposite signs is important for the physics of the Lifshitz transition, and is a consequence of the fact that $\De$ pairs opposite valleys, and hence anticommutes with the $\tau^z$ contained in $\bfsig_v$. Note that this physics is independent of our assumption of even angular momentum pairing, since it only relies on the fact that $\De$ is off-diagonal in the valley index. 
		
		Performing the trace, we then find 
		\bea 
		\mcl_2[\De(0,&q_x/ v_x,q_y/v_y)] \supset - \frac{|\De|^2}{v_xv_y} T \sum_n \int_{\bfk}  \frac1{(\o^2_n+k^2)(\o_n^2+(\bfk+\bfq)^2)} \( \o_n^2 (\l+h\G)^2 +(\bfk+\bfq)\cdot\bfk (\l^2 +h\G)^2  \)
		\eea 
		Introducing a Feynman parameter $x$ (integrated from 0 to 1), 
		\bea 
		\mcl_2[\De(0,&q_x/ v_x,q_y/v_y)]   \supset - \frac{|\De|^2(\l+h \G)^2}{v_xv_y} T \sum_n \int_{\bfk,x} \frac{\o_n^2 + k^2 - q^2y}{(\o_n^2 + k^2 + q^2y)^2} \\
		\eea 
		where $y=x(1-x)$.

		Doing the sum over Matsubara frequencies,
		\bea\label{second_o_matsu} \mcl_2[\De(0,&q_x/ v_x,q_y/v_y)] 
		\supset - \frac{|\De|^2(\l+h \G)^2}{2Tv_xv_y} \int_{\bfk,x} \Bigg( \frac{q^2y \, \sech^2(\sqrt{k^2+q^2y}/2T)}{k^2+q^2y} + \frac{2k^2 T}{(k^2+q^2y)^{3/2}} \tanh(\sqrt{k^2+q^2y}/2T) \Bigg). \eea 
		The exact expression for this integral is not so important for the present purposes, as we are mostly just interested in the signs of the order $q^0,q^2$ terms. We find 
		\bea \mcl_2[\De(0,&q_x/ v_x,q_y/v_y)] 
		\supset  \frac{|\De|^2(\l+h\G)^2}{2v_xv_y}(-C_1 + C_2 q^2),\eea
		where $C_1,C_2$ are positive (temperature-dependent) coefficients. 
		Selecting out the contribution which depends on the displacement and magnetic fields $U$ and $A$ by using the definition of $\G$, and dropping terms of order $A^4$, we then have 
		\bea \label{quadratic_terms} \mcl_2[\De(0,\bfq)] \supset  \frac{2\l h(U^2/4 - v_D^2 A^2) + h^2(U^4/16 - U^2 v_D^2 A^2 /2)}{2v_xv_y}\( -C_1  |\De|^2 + C_2 (v_x^2 |\p_x\De|^2 + v_y^2|\p_y \De|^2)\).\eea

		\sss{Quartic terms}
		
		We now examine how the orbital field affects the the quartic terms $u,v$ in the GL free energy.
		Using $\cpl^\da \De \cpl^* = \G \De$, we have (now restricting to zero momentum transfer)
		\bea \label{quartic_terms} \mcl_4[\De]  & \supset +T\isum \frac{\Tr[((h\cpl^\da \De\cpl^* + \l \De) (i\o_n + \bfk\cdot\bfsig_v)(h\cpl^T \De^\da\cpl + \l \De^\da) (i\o_n + \bfk\cdot\bfsig_v^*))^2]}{(\o_n^2+k^2)^4}\\
		& =T(h\G+\l)^4\isum \frac{\Tr[(\De\De^\da)^2]}{(\o_n^2+k^2)^2} \\ 
		& =(h\G+\l)^4 \frac{\Tr[(\De\De^\da)^2]}{16\pi T} \\
		& =  (h\G+\l)^4\frac{ 2|\De|^4 - \De^2 (\De^*)^2}{16\pi T}.
		\eea 
		
		\sss{Summary}

		To understand the consequences of the terms derived so far, consider the simplified free energy $f_s = \frac c2|\p \De|^2 + r|\De|^2 + \frac u2|\De|^4$, where the phase stiffness $K$ is related to $c$ as $K = |\De|^2 c$. The effects derived in  \eqref{quadratic_terms}, \eqref{quartic_terms}  lead to renormalizations of $K,r,u$ by amounts 
		\bea \label{renorms} \d r & = - C_1\frac{2\l h(U^2/4 -v_D^2A^2) + h^2(U^4/16 - U^2 v_D^2A^2/2)}{2v^2} \\ 
		\d K & = 2|\De|^2C_2\frac{2\l h(U^2/4 -v_D^2A^2) + h^2(U^4/16 - U^2 v_D^2A^2/2)}{2v^2}  \\ 
		\d u & = \frac{(h(U^2/4 - v_D^2A^2) + \l)^4}{4\pi T}\eea 
		where we have assumed $v_x=v_y\equiv v$ for simplicity. 
		
		The condensate fraction $\lan |\De|^2\ran = -r/u$ is renormalized by 
		\be \d \lan |\De|^2\ran = -\d r\frac{1}{u_0} + \d u\frac{r_0}{u_0^2}, \ee 
		where $r_0,u_0$ are the values of $r,u$ in the absence of any coupling between the mirror-even and mirror-odd fermions. 
		
		The exact field dependence of $\d \lan |\De|^2\ran$ depends on the magnitude of $\l$ and the sign of $h\l$, which are not things that can be derived microscopically within the current approach. First consider the case where $\l$ is small enough so that terms of order $\l h$ may be dropped compared to those of order $h^2$, and the contribution from the $\d u$ term can be ignored. This very well may be the case microscopically, as the interaction which produces the $\l$ term does not appear in the (presumably dominant) density-density interaction (see the discussion following \eqref{masterl}). In this case we have \be \d \lan |\De|^2 \ran \approx \frac{C_1h^2}{2v^2u_0}(U^4/16 - U^2 v_D^2A^2/2).\ee 
		Therefore the displacement field increases the condensate fraction, while the magnetic field decreases it. The same dependence also holds for $K$. Note in particular that in this limit, the condensate fraction is unchanged by the magnetic field when $U=0$ (at least to order $B^2$), implying that to this order the critical in-plane magnetic field $B_c$ is infinite.
		
		When $\l h$ is not small compared to $h^2$, it is possible to engineer a situation where the behavior of the condensate fraction is reversed from the discussion above (provided that $\l h<0$). This however can be ruled out from experiment, given that the condensate fraction is seen to increase with $U$ at $B=0$ (at least for the range of displacement fields we are interested in). Therefore in this case the leading shifts of the condensate fraction and of $K$ are both proportional to $U^2/4-v_D^2A^2$, with positive constants of proportionality. Note that this gives a finite value of $B_c$ even when $U=0$, as in experiment.  This could either be because  the experimental system  has a nonzero (though possibly small) value of $\l$, or because the mirror symmetry is not perfect.  
		
		Summarizing, we conclude that the displacement field and (orbital) magnetic field have {\it opposite} effects: the displacement field $U$ {\it increases} both the condensate fraction and the superfluid stiffness, while the magnetic field {\it decreases} both the condensate fraction and the superfluid stiffness. 
		Because $K$ and $r$ are renormalized by different amounts, it is reasonable to consider a scenario where the superfluid stiffness goes to zero with increasing $B$ before the condensate fraction does.

		\ss{Zeeman effects \label{ss:zeeman} }
		
		We now turn to examining contributions to the effective action for $\De$ arising from the Zeeman coupling of the magnetic field to the microscopic fermions. 
		These contributions come from the term 
		\be \mcl_Z = -\sfg \(  \psi_+^\da  \bfB \cdot \bfs \psi_+ + \psi_2^\da \bfB \cdot \bfs \psi_2 +   \psi_-^\da \bfB \cdot\bfs \psi_-\),\ee 
		where for brevity we have defined $\sfg \equiv \mu_B g / 2$. 
		
		\sss{Response from a Fermi liquid} 
		Modeling the mirror-even fermions as a Fermi liquid as described at the beginning of this appendix, we can 
		incorporate the Zeeman effects they induce by using the propagators 
		\bea G_p & = \frac{i\o-\xi_\bfk -\gplus \bfB\cdot\bfs }{-(-i\o+\xi_\bfk)^2+\gplus^2B^2} 
		\\ G_h &= \frac{i\o+\xi_\bfk +\gplus \bfB\cdot\bfs^* }{-(-i\o-\xi_\bfk)^2+\gplus^2B^2} ,
		\eea 
		where $\xi_\bfk$ is the energy measured relative to the Fermi energy. 
		This then provides the term 
		\bea \label{quad_zeeman} \mcl_2 [\De(0,\bfq)] & \supset \Tr[G_p \De G_h \De^\da ] \\ 
		& = T\sum_{\o_n} \int_\bfk \frac{\Tr[(i\o_n -\xi_\bfk - \gplus  \bfB\cdot\bfs)\De (i\o_n + \xi_{\bfk+\bfq } + \gplus  \bfB\cdot\bfs^*)\De^\da]}{(-(-i\o_n + \xi_\bfk)^2 + \gplus^2B^2)(-(-i\o_n - \xi_{\bfk+\bfq})^2+\gplus^2B^2)}\eea 
		
		At linear order in $\bfB$, this gives the contribution 
		\bea \mcl_2[\De(0,\bfq)] & \supset  
		\gplus  T\sum_{\o_n} \int_\bfk \Tr\[ \frac{\bfB\cdot\bfs \De \De^\da }{(-i\o_n+\xi_\bfk)^2(-i\o_n - \xi_{\bfk+\bfq})} - \frac{\De\bfB\cdot\bfs^* \De^\da }{(-i\o_n+\xi_\bfk)(-i\o_n - \xi_{\bfk+\bfq})^2}\] \\ 
		& = \gplus  T\sum_{\o_n}\int_\bfk \frac{\Tr[ \bfB \cdot\bfs \De \De^\da + \De \bfB\cdot\bfs^* \De^\da]}{(-i\o_n + \xi_{\bfk})^2(-i\o_n - \xi_{\bfk+\bfq})}
		\eea 
		where we shifted the momentum integration and used that the response will only be a function of $q^2$, so that $\xi_{\bfk-\bfq}$ can be replaced with $\xi_{\bfk+\bfq}$.
		Evaluating the trace, we find 
		\bea \mcl_1[\De(0,\bfq)] & = -8\gplus  \bfB \cdot  \bfM \, I(q),\eea 
		where as before $\bfM = \bfd_1\times \bfd_2$, and where
		\bea I(q) & = T \sum_{\o_n} \int_\bfk \frac{1}{(-i\o_n + \xi_{\bfk})^2(-i\o_n - \xi_{\bfk+\bfq})}
		\eea 
		For simplicity we will only explicitly evaluate the $O(q^0)$ term. In the limit where we can replace $\int_\bfk$ with $N(0)\int_{-\L}^\L d\xi$, this term vanishes due to particle-hole symmetry about the Fermi level. The leading contribution is thus proportional to $N'(0)$ \cite{sugiyama1995mechanism}, giving 
		\be I(0) \approx - \frac{N'(0)}4 \int \frac{d\xi}\xi \tanh(\xi/2T) \approx \frac{N'(0)}4\ln \( \frac{\L}{T}\).\ee 
		Due to the smallness of $N'(0)$ in a generic Fermi liquid scenario, this term would normally be neglected. However, it is possible that the displacement field modifies the density of states significantly enough to make this term important at moderate $U$. 
		
		Before moving on, we note that the above $\bfB \cdot \bfM$ term was derived by performing an expansion in powers of $\De$, using the normal state form of the fermion propagators. In the limit where $T\ll\De$, we should instead perform an expansion about the $B=0$ superconductor. In this case the $\bfB\cdot\bfM$ term simply vanishes, as there is no magnetization for the field to couple to.

		\ms 
		
		Now we look at effects from \eqref{quad_zeeman} which enter at order $B^2$. 
		Using 
		\bea \Tr[\bfB \cdot \bfs \De \bfB \cdot\bfs^* \De^\da] 
		=-2|\bfB\cdot\bfd|^2 + B^2(|\bfd|^2 - |d_0|^2),
		\eea 
		the contribution at order $B^2$ is 
		\bea  \label{fl_bsq} \mcl_2[\De(0,\bfq)] & \supset \gplus^2 T \isum \Bigg[ \frac{2|\bfB\cdot\bfd|^2-B^2(|\bfd|^2-|d_0|^2)}{(-i\o_n+\xi_\bfk)^2(-i\o_n-\xi_{\bfk+\bfq})^2} \\ & \qq\qq + \frac{B^2|\De|^2}{(-i\o_n +\xi_{\bfk})(-i\o_n -\xi_{\bfk+\bfq})} \( \frac1{(-i\o_n+\xi_{\bfk})^2} + \frac1{(-i\o_n-\xi_{\bfk+\bfq})^2}\)\Bigg]\\ 
		& = \gplus^2 T \isum \Bigg[ \frac{2|\bfB\cdot\bfd|^2-B^2(|\bfd|^2-|d_0|^2)}{(-i\o_n+\xi_\bfk)^2(-i\o_n-\xi_{\bfk+\bfq})^2} + \frac{2B^2 |\De|^2}{(-i\o_n +\xi_{\bfk})^3(-i\o_n -\xi_{\bfk+\bfq})}\Bigg]  \eea 
		where the second term in the first line comes from the expansion of the $B^2$s in the denominators of the propagators, and we have again used that the response is only a function of $q^2$ when going to the second line. 
		
		The general expressions that appear after doing the frequency sum are rather complicated. For our purposes it suffices to expand in $q^2$, and write 
		\be   \mcl_2[\De(0,\bfq)]  \supset \gplus^2 \( |\bfB\cdot\bfd|^2 (F_1 - F_2 q^2) + B^2|\bfd|^2 (G_1 - G_2 q^2) + B^2 |d_0|^2(H_1 -H_2q^2)\),\ee 
		where the $F_i,G_i,H_i$ are all positive functions of $T$, with all but $F_1,H_1$ vanishing in the limit where we may perform the $\bfk$ integrals as $N(0) \int_{-\infty}^\infty d\xi$.

		\sss{Response from a Dirac cone} 
		
		We now compute the Zeeman effects induced by the mirror-odd Dirac fermion $\psi_-$ (these are expected to be smaller, but we include the discussion for completeness). 
		To do this, we need to compute the propagator for $\psi_-$ in the absence of hybridization with $\De$. To this end, consider the matrix 
		\be M = a \unit + \bfb \cdot \bfs + \bfc \cdot \bfsig_v.\ee 
		One can show that the inverse is 
		\be M\inv = \frac{a(a^2-b^2-c^2) + (b^2-a^2-c^2) \bfb\cdot\bfs + (c^2-a^2-b^2) \bfc\cdot\bfsig_v + 2a\bfb\cdot\bfs\tp\bfc\cdot\bfsig_v}{a^2(a^2-b^2-c^2) + b^2(b^2-a^2-c^2) + c^2(c^2-a^2-b^2)}.\ee 
		If we apply this to the case of $a = -i\o-\mu, \bfb= -\gminus  \bfB, \bfc  = \bfk$, we expand to order $B^2$ and get 
		\bea G_p & = \frac{i\o_- + \bfk\cdot\bfsig_v}{\o_-^2+k^2} + \gminus  \bfB \cdot \bfs \frac{-\o_-^2 +k^2+2i\o_- \bfk\cdot\bfsig_v}{(\o_-^2+k^2)^2} - \gminus^2B^2 \frac{i\o_-^3 - 3i\o_- k^2+(3\o_-^2-k^2)\bfk\cdot\bfsig_v}{(\o_-^2+k^2)^3}\\
		G_h & = \frac{i\o_+ + \bfk\cdot\bfsig_v^*}{\o_+^2+k^2} - \gminus  \bfB \cdot \bfs^* \frac{-\o_+^2 +k^2+2i\o_+ \bfk\cdot\bfsig_v^*}{(\o_+^2+k^2)^2} - \gminus^2B^2 \frac{i\o_+^3 - 3i\o_+ k^2+(3\o_+^2-k^2)\bfk\cdot\bfsig_v^*}{(\o_+^2+k^2)^3},\eea 
		where $\o_\pm = \o \pm i\mu$. 
		We then use these propagators to integrate out $\psi_-$ and compute the resulting terms in the effective action for $\De$, following the same approach as when considering orbital effects. 
		
		The contribution linear in $\gminus B$ vanishes if $\mu=0$ due to particle-hole symmetry, but is nonzero otherwise. To leading (linear) order in $\mu$,
		\bea  \mcl_2[\De(0,\bfq)] & = \gminus  (\l +h\G)^2 T \isum \Bigg[ \frac{\Tr[\bfB\cdot\bfs (-\o_-^2 + k^2 + 2i\o_- \bfk\cdot\bfsig_v ) \De (i\o_+ + \bfk\cdot\bfsig_v^*) \De^\da ]}{(\o_-^2+k^2)^2(\o_+^2+k^2)} \\ & \qq\qq\qq - \frac{\Tr[(i\o_- + \bfk\cdot\bfsig_v )\De \bfB \cdot\bfs^* (-\o_+^2 +k^2+2i\o_+\bfk\cdot\bfsig^*)\De^\da ]}{(\o_-^2 + k^2)(\o_+^2 + k^2)^2}\Bigg] \\ 
		& = -L\mu(\l +h\G)^2 \gminus    \Tr[\bfB\cdot\bfs \De \De^\da + \De \bfB \cdot\bfs^* \De^\da ] \\ 
		& = 4L\mu (\l +h\G)^2 \gminus  \bfB \cdot\bfM,
		\eea 
		where $L$ is a positive constant.

		Contributions at order $(\gminus B)^2$ give the term (now dropping the chemical potential for simplicity) 
		\bea  \label{dirac_bsq} \mcl_2[\De(0,\bfq)] & \supset - \gminus^2 (\l +h\G)^2T \isum \Bigg[ \frac{\Tr[\bfB\cdot\bfs (-\o_n^2+(\bfk+\bfq)^2+2i\o_n (\bfk+\bfq)\cdot\bfsig_v )\De \bfB\cdot\bfs^* (-\o_n^2 + k^2+ 2i\o_n \bfk\cdot\bfsig_v^* )\De^\da]}{(\o_n^2+(\bfk+\bfq)^2)^2(\o_n^2+k^2)^2} \\ 
		& \qq\qq + 2B^2 \frac{\Tr[(i\o_n+(\bfk+\bfq)\cdot\bfsig_v )\De (i\o_n^3-3i\o_n k^2 + (3\o_n^2-k^2) \bfk\cdot\bfsig^*_v)\De^\da ]}{(\o_n^2+(\bfk+\bfq)^2)(\o_n^2+k^2)^3} \Bigg] \\ 
		& = -\gminus^2(\l +h\G)^2 T\isum\Bigg[ \frac{-2|\bfB\cdot\bfd|^2 + B^2(|\bfd|^2-|d_0|^2)}{(\o_n^2+(\bfk+\bfq)^2)^2(\o_n^2+k^2)^2}\( (-\o_n^2-(\bfk+\bfq)^2)(-\o_n^2+k^2) + 4\o_n^2 (\bfk+\bfq)\cdot\bfk \) \\ 
		& \qq\qq - 2B^2 |\De|^2\( \frac{\o_n^4 -3k^2\o_n^2+ (\bfk+\bfq)\cdot\bfk (3\o_n^2-k^2)}{(\o_n^2+(\bfk+\bfq)^2)(\o_n^2+k^2)^3} \) \Bigg] 
		\eea 
		Evaluating the integrals gives %
		\be  \mcl_2[\De(0,\bfq)]  \supset  (\l +h\G)^2\gminus^2\( |\bfB\cdot\bfd|^2 (P_1 -P_2q^2) - B^2|\bfd|^2(Q_1 - Q_2q^2) + B^2 |d_0|^2(R_1-R_2q^2) \),\ee 
		where as usual $P_i,Q_i,R_i$ are positive (temperature dependent) constants.

		\section{Details on the Ginzburg-Landau theory \label{app:lg_details}}
		
		In this section we detail the phases described by the Ginzburg-Landau theory discussed in the main text. See \cite{scheurer2020pairing} for a similar analysis in a more general context. 
		
		\ss{Fully spin-valley locked limit \label{app:svl_lg}} 
		
		We will first consider the case where the spin is fully locked to valley, so that the order parameter is allowed to be nonzero only in a subset of the Hilbert space. 
		
		As was discussed above near \eqref{2mop}, in the case of full SVL, we have $\det(\De_{KK'}) = 0$ and may
		write the order parameter as 
		\be \De = \frac{|\De|}2  \bpm & -|\eta\ran \lan \xi^*|\\ 
		|\xi\ran \lan \eta^* | \epm,\ee 
		where as before $|\De| = \sqrt{\Tr[\De^\da\De]}$, the matrix written out explicitly is in valley space, and where $\xi, \eta$ are normalized spinors which determine the directions along which the spins in the $K,K'$ valleys are polarized.
		
		By decomposing the outer products appearing in $\De$ in terms of Pauli matrices, we may write (assuming even angular momentum pairing for concreteness, as in the main text)
		\bea \De & = \frac{|\De|}4 \( i\tau^y ( \bfs \,is^y) \cdot   \lan\ob \xi|  \bfs |\eta\ran + \tau^x is^y \lan \ob\xi |\eta\ran\)\\ 
		& = \frac{|\De|}{2^{3/2}}\(i\tau_y \bfd \cdot \bfs  + \tau^x d_0\) is^y 
		,\eea 
		where the $d$ four-vector is {\it defined} in terms of $\k\xi,\k\eta$, and has components 
		\be d_\mu = \frac1{\sqrt2}\lan \ob \xi | s^\mu |\eta\ran,\ee 
		where we have defined the adjoint spinor\footnote{The adjoint of a given spinor is obtained by inverting the spinor through the Bloch sphere and changing the handedness used to define its overall phase.
			In terms of the Euler angle parametrization 
			\be \k\xi = e^{i\g/2}\bpm e^{i\phi/2} \cos(\t/2) \\ e^{-i\phi/2} \sin(\t/2) \epm,\ee 
			the adjoint $\k{\ob\xi}$ can be obtained by by sending $\g\mt -\g +\pi,\,  \phi \mt \phi +\pi,\, \t \mt -\t + \pi.$ }
		\be \label{adjoint_def} \k{\ob \xi} = -i s^y  \k\xi^*.\ee

		In what follows it will be helpful to define, for any spinor $\k\vs$, the three-vector 
		\be \boldsymbol{\vs} = \lan \vs |\bfs |\vs\ran \implies | \vs \ran \lan \vs | = \frac12(1 + \boldsymbol{\vs}\cdot\bfs ).\ee 
		Taking the adjoint in the manner above simply reverses the direction of the 3-vector, so that $\boldsymbol{\ob\vs} = - \boldsymbol{\vs}$. This notation lets us write the magnitudes of $\bfd$ and $d_0$ as 
		\bea |\bfd|^2 & = \frac12 \lan \ob\xi |s^j |\eta\ran\lan \eta| s^j | \ob\xi\ran 
		\\
		& = \frac14(3+\bfeta \cdot \bfxi), 
		\\ |d_0|^2 & = \frac14| \lan \ob\xi|\eta\ran|^2
		\\
		& = \frac14(1 -\bfxi\cdot\bfeta).
		\eea 
		The norm of the full $d$ vector is thus $|d_\mu|^2 = 1$, giving $\Tr[\De^\da \De] = |\De|^2$ as required. Also as expected for a spin-polarized state, we see that the weight of the triplet part $|\bfd|^2$ is always nonzero, being minimal for antiferromagnetic alignment between the valleys and maximal for ferromagnetic alignment, with the opposite being true for the weight of the singlet part $|d_0|^2$.

		When deriving the coefficients in the GL free energy microscopically, SVL can be taken into account simply by replacing the particle and hole propagators by  
		\be \label{proj_props} G_p \mt\mcp G_p \mcp,\qq G_h \mt \mcp^T G_h \mcp^T,\ee 
		where we have defined the projector 
		\be \mcp = \kb{\eta} \oplus \kb{\xi},\ee 
		with the $\oplus$ in valley space. Note that $\mcp$ satisfies 
		\be \mcp \De = \De ,\qq \De \mcp^T = \De.\ee 
		
		The form of terms not involving the Zeeman coupling of the magnetic field are not modified by the projection. For terms involving the Zeeman coupling, we simply make the replacement 
		\be \bfB \cdot\bfs \mt \mcp \bfB \cdot \bfs \mcp = \bfB \cdot \bpm \bfeta \kb\eta & \\ & \bfxi \kb\xi \epm  \equiv \mcp_B.\ee 
		
		The structure of the Zeeman terms quadratic in the magnetic field are determined by following the analysis of appendix \ref{app:lg_coeffs}, and using the traces
		\bea \Tr[\mcp_B \De \De^\da + \De \mcp_B^* \De^\da ] & = 2 |\De|^2 \bfB\cdot(\bfeta + \bfxi),\\ 
		\Tr[\De^\da \mcp_B^2 \De + \De^\da \De (\mcp_B^*)^2] & = 2 |\De|^2 ((\bfB\cdot\bfeta)^2 +  (\bfB \cdot\bfxi)^2 )\\
		\Tr[\De^\da \mcp_B \De \mcp_B^* ]  & = 2|\De|^2 (\bfB\cdot\bfeta)(\bfB\cdot\bfxi).  \eea 
		For example, for a spin-valley locked Fermi liquid, the contributions to the free energy from the Zeeman couplings are, to quadratic order in $\De$ and $B$, 
		\be\z |\De|^2\bfB \cdot( \bfxi + \bfeta) + \kappa |\De|^2(\bfB \cdot (\bfxi -\bfeta))^2,\ee 
		where $\z$ is small in $N'(0)/N(0)$ and $\kappa>0$ (see appendix \ref{app:lg_coeffs}). As expected, we may avoid Pauli suppression of the pairing either by having the spins in each valley align in the plane normal to $\bfB$, or by having the two spins ferromagnetically align along $\bfB$, both of which prevent pair-breaking by the field from occurring. 
		
		The free energy in general may be written as a function of $|\De|,\bfxi,\bfeta$ as (ignoring derivatives of $\bfxi,\bfeta$ for simplicity)
		\bea f & = K |\p \De|^2 + r|\De|^2 + \frac{u}2|\De|^4 +  w |\De|^2\bfeta \cdot \bfxi  + \kappa|\De|^2 (\bfB\cdot(\bfxi-\bfeta))^2- \zeta  |\De|^2 \bfB \cdot(\bfxi+\bfeta)\eea 
		where the coefficient $w$ determines whether the spins in each valley prefer to align ferromagnetically or antiferromagnetically. 
		The explicit $\bfB$ dependence arises from the Zeeman coupling terms, while the orbital effects of the field provide implicit dependence of the various coefficients on $B^2$ in the manner described in appendix \ref{app:lg_coeffs}.

		As discussed in the main text, the system observed in experiments likely has $w>0$. When $B=0$, the minimum of $f$ occurs for $\bfxi = -\bfeta$. $\De$ is now invariant under $SO(2)$ rotations about $\bfxi$, and hence the order parameter manifold is $(U(1)\times S^2)/\zt$. 
		
		Now we will examine what happens to this state at nonzero $\bfB$. 
		Taking $\bfB$ to lie along $\uvz$ in spin space, upon minimizing $f$ we find
		\be \bfxi = R_{z}(\vp) \bpm 0 \\ \sin \t \\ \cos\t \epm ,\qq \bfeta = R_z(\vp) \bpm 0 \\ - \sin \t \\ \cos\t \epm,\ee 
		where $R_z(\vp)$ represents a rotation by an arbitrary angle $\vp$ about $\uvz$, and where
		\be \t = \begin{dcases} \arccos(|\z B| / 2w),\qq & |\z B|/2w \leq  1 \\ 
			0 ,\qq & |\z B|/2w > 1 \end{dcases}. \ee 
		Thus for $0 < |\z B|/2w < 1$ the spins are partially magnetized and there is an $U(1)$s worth of degenerate minima, and as in the partially polarized phase of the following subsection the order parameter manifold is $(U(1)\times U(1))/\zt$. For $|\z B|/2w >1$ the spins are ferromagnetically aligned and fully magnetized, and the order parameter manifold is $U(1)$.
		Given this solution for $\bfxi,\bfeta$, the magnitude of the order parameter is consequently determined to be 
		\be \label{full_cond_frac} \lan |\De|\ran = \sqrt{-\wt r/u},\qq \wt r = r + w(2\cos^2(\t) -1) - 2|\z B| \cos(\t).\ee 
		The condensate fraction and its first derivative with respect to $B$ are continuous across the transition into the high-field ferromagnetic state, and $(\p \lan |\De|\ran / \p B)|_{B=0} = 0$, implying that unlike in the non-locked case to be discussed in the next subsection, $T_{SC}$ will not increase linearly with $B$ at small $B$. 
		
		\begin{figure}
			\subfloat[]{%
				\includegraphics[width=.33\textwidth]{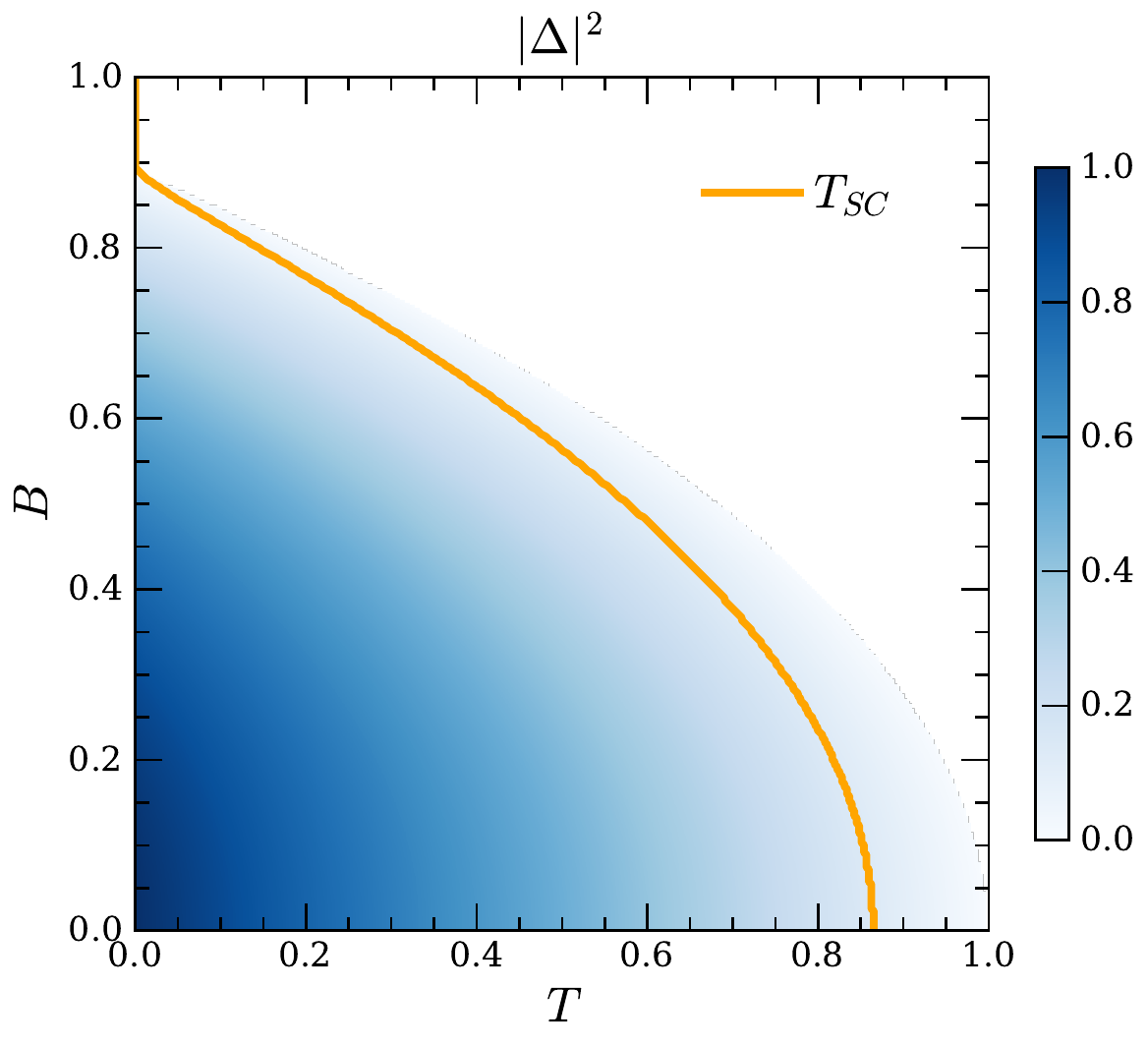}%
			}\hfill
			\subfloat[]{%
				\includegraphics[width=.33\textwidth]{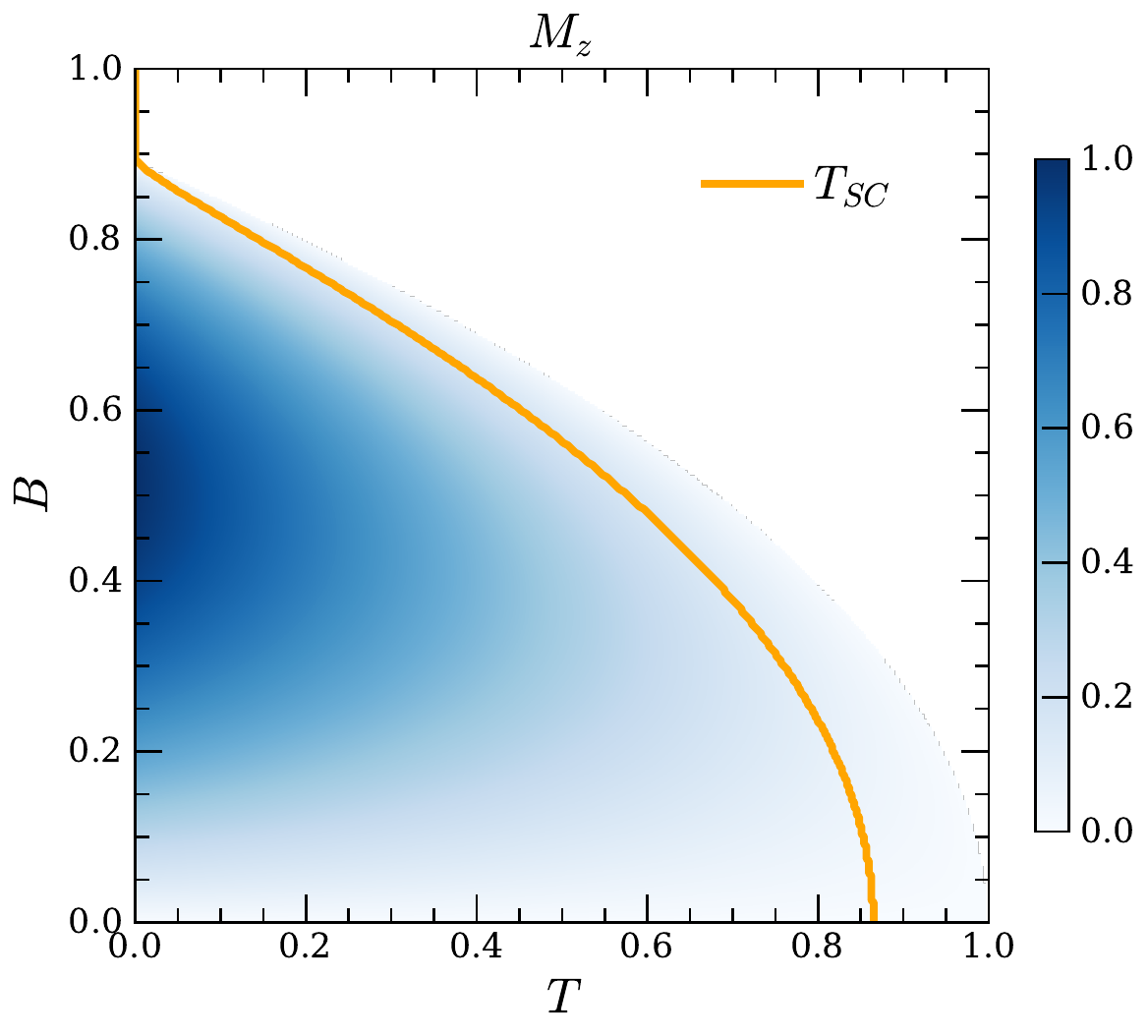}%
			}
			\hfill 
			\subfloat[]{%
				\includegraphics[width=.33\textwidth]{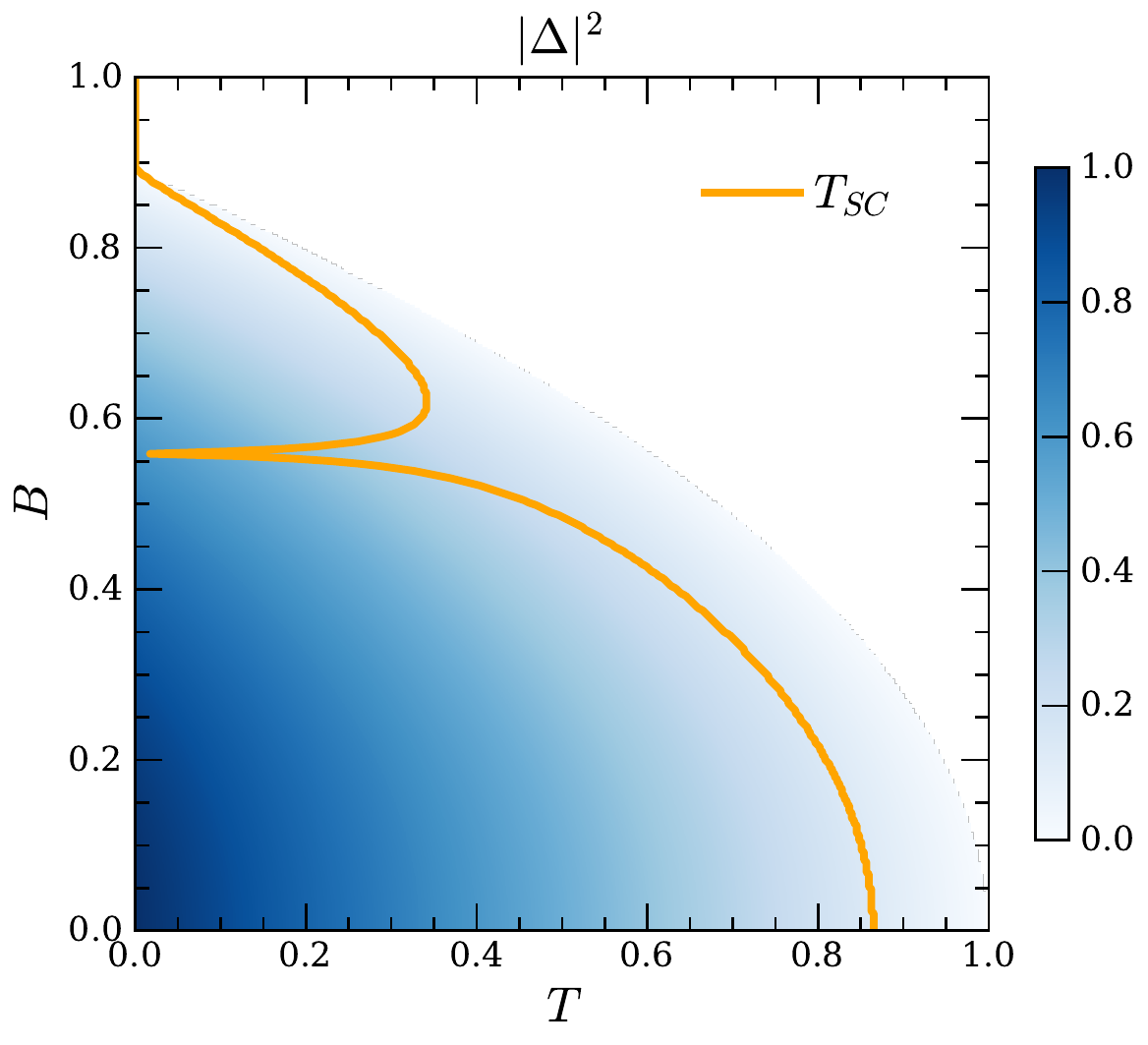}%
			}
			\caption{a) condensate fraction in the spin-valley locked case (normalized by $|\De(B=0,T=0)|^2$) for a typical choice of parameters, and ignoring dependence of $K$ on $B$. $T_{SC}$ marks the BKT transition temperature for the superconductor. b) expected value of the component of the magnetization along the field direction, normalized to $M_z(B=0,T=0)$. c) the same as a), but taking into account a downward renormalization of $K$ by a term proportional to $B^2$, leading to a field-induced Lifshitz transition. Here we have taken $\z < 2w/B_c$, where $B_c$ is the critical field for the superconductor; as such the system never reaches the fully magnetized regime. }
			\label{fig:svl_cond_fracs}
		\end{figure}
		
		Figure \ref{fig:svl_cond_fracs} shows the condensate fraction (left panel) and magnetization (center panel) for a typical choice of parameters (here $\z$ is small enough so that the system never reaches the fully magnetized regime). The right panel is the same as the left, but shows the BKT temperature after taking into account orbital effects which suppress $K$.

		\ss{No spin-valley locking} 
		
		Now we consider the slightly simpler case in which there is no spin-valley locking (SVL), with the order parameter allowed to have components in the entire fermionic Hilbert space. This discussion will therefore only be germane for the problem considered in this paper at values of $T$ and / or $B$ large enough to eliminate any SVL which occurs in the $T=0$ normal state near $\nu = -2$. 
		
		We start by breaking the order parameter up into triplet and singlet parts as 
		\bea \De & =  i\tau^y\, \bfd \cdot \bfs\, is^y  + \tau^x \, d_0\, is^y,\eea 
		where $d_\mu = (d_0,\bfd)$ is a complex four-vector. Up to quartic order in the order parameter, the free energy can be written as 
		\bea \label{app_full_lg} f & = f_t + f_s + f_{ts} \\ 
		f_t & = \sum_{i=1,2} \(K(\p \bfd_i)^2 + r d_i^2 + \frac u2d_i^4 + \kappa(\bfB\cdot \bfd_i)^2\) +  (u+v) d_1^2d_2^2 - v (\bfd_1\cdot\bfd_2)^2 + \z \bfB\cdot(\bfd_1 \times \bfd_2)  \\ 
		f_s & = K' (\p d_0)^2 + r' |d_0|^2 + \frac{u'}2 |d_0|^2 + \kappa'B^2 |d_0|^2   \\ 
		f_{ts} & = q |d_0|^2 (d_1^2+d_2^2) + w |d_0^2- \bfd^2|^2 \eea
		where we have used the standard notation $\bfd = \bfd_1 + i \bfd_2$. As usual, the terms which depend explicitly on the magnetic field arise from Zeeman effects, while orbital effects are responsible for giving the remaining coefficients implicit dependence on $B^2$. 
		
		When minimizing $f$ in the presence of of the couplings to the magnetic field, it is often helpful to write the $\bfd$ vector in terms of components which diagonalize the $\bfB\cdot(\bfd_1 \times \bfd_2)$ coupling \cite{kawaguchi2012spinor}. Taking the magnetic field to point in the $\uvz$ direction in spin space, this is done through the unitary transformation
		\be \bpm d_x\\ d_y \\ d_z \epm = \frac1{\sqrt2} \bpm 1 & & 1 \\ i & & -i\\ &\sqrt2& \epm \bpm \duu \\ \dud \\ \ddd \epm, \ee 
		In this representation, the $f_t$ and $f_{ts}$ components of the free energy are now 
		\bea f_t & = K\sum_{\s=+,-,z} ( |\D d_\s|^2 + r  |d_\s|^2)   + \frac{u+v/2}2( |\duu|^2 + |\ddd|^2 + |\dud|^2)^2  - v |\duu \ddd +\dud^2/2|^2 \\ & \qq\qq + \kappa B^2 |\dud|^2 - \frac{\z B}2 (|\duu|^2 - |\ddd|^2) \\ 
		f_{ts} & = q|d_0|^2( |\duu|^2 + |\ddd|^2 + |\dud|^2) + p|d_0^2 - (2\duu \ddd + \dud^2)|^2.
		\eea

		The effects of the $p$ term can be understood as a way of softly implementing SVL. Indeed, if we write the order parameter as 
		\be \label{2mop} \De = \bpm & -\De_{KK'}^T \\ \De_{KK'}\epm, \ee 
		(where the 2$\times2$ matrix is in valley space) we see that we may write 
		\be |d_0^2 - \bfd^2|^2 = |d_0^2 - (\dud^2+2\duu \ddd)|^2 = |\det(\De_{KK'})|^2.\ee 
		Therefore large positive $p$ enforces that the determinant of $\De_{KK'}$ vanish. In this case we may write 
		\be \De_{KK'} \propto |\xi \ran \lan \eta^*| \ee 
		for two spinors $\k\xi,\k\eta$, which have the interpretation as the directions along which the spins in the $K,K'$ valleys are polarized. 
		For simplicity of the analytical results, and because the fully spin-valley locked case was discussed in the last subsection, in what follows we will simply set $p=0$. 
		
		Extremizing the free energy for uniform solutions then gives 
		\bea\label{free_energy_min} r\duu + (u+v/2)\duu d^2 - v \ddd^*(\duu\ddd +\dud^2/2)- \z B \duu/2 + q|d_0|^2 d_+ & = 0  \\ 
		r\ddd + (u+v/2) \ddd d^2 - v \duu^*(\duu\ddd +\dud^2/2) + \z B\ddd/2 + q|d_0|^2 d_-& = 0 \\ 
		r\dud + (u+v/2)\dud d^2 + v\dud^*(\duu\ddd +\dud^2/2) + \kappa B^2 \dud + q|d_0|^2 d_z& = 0   \\ 
		(r'+\kappa'B^2 + u'|d_0|^2 +  q( |\duu|^2 + |\ddd|^2 + |\dud|^2))d_0 & = 0 
		\eea 
		
		One can show that when $\kappa>0$, which is the case we are interested in, when $B\neq 0$ the global minima of the free energy always have $d_z = 0$, while when $B=0$ we have $SO(3)$ spin symmetry and may fix $d_z=0$ without loss of generality. Setting $d_z=0$ then, when $d_0\neq0$ we just have to solve 
		\bea \wt r\duu + (\wt u+v/2)\duu d^2 - v \duu|\ddd|^2- \z B \duu/2 & = 0  \\ 
		\wt r\ddd + (\wt u+v/2) \ddd d^2 - v \ddd|\duu|^2  + \z B\ddd/2 & = 0 \eea
		where 
		\be \wt r = r - \frac{q}{u'}(r'+\kappa'B^2),\qq \wt u = u - q^2/u'\ee
		and\footnote{When the RHS of \eqref{dssq} becomes negative, the minimum of $f$ has $d_0=0$. The solutions in this case are obtained simply by replacing $\wt r, \wt u$ with $r,u$. }
		\be \label{dssq} |d_0|^2 = \frac{-r' -\kappa'B^2 - q(|d_+|^2 + |d_-|^2)}{u'}.\ee 
		There are now two types of solutions. The first is with $d_+=0$ or $d_-=0$, corresponding to fully spin-polarized triplets. Choosing $\z B>0$, the favored solution will be the one with $d_-=0$. This gives the fully polarized solution 
		\be \label{fully_polarized} \bfd_{fp} = \bpm e^{i(\phi +\g)}\displaystyle\sqrt{\frac{ \z B - 2\wt r}{2\wt u+v}},&0,&0\epm,\ee 
		where $\phi$ ($\g$) shifts under $U(1)_g$ gauge transformations ($SO(2)_s$ spin rotations about $\uvz$). 
		This order parameter is therefore invariant under combined $U(1)_g$ and $SO(2)_s$ rotations. As a result, when $B=0$ the order parameter manifold is $\mcm = (U(1)_g \times SO(3)_s) / U(1)_{g+s} = SO(3)$ \cite{cornfeld2020spin}, since the $U(1)_g$ subgroup can be absorbed into the $SO(3)$ part. As discussed in the main text, this means that at $B=0$ there are only $\zt$ vortices and that $T_{BKT} = 0$, which is incompatible with experiment. When $B\neq0$ the order parameter manifold is instead $\mcm = (U(1)_g \times U(1)_s)/ U(1)_{g+s} = U(1)$, so that $\zz$ vortices (and a nonzero $T_{BKT}$) occur at finite fields. 
		
		When both $d_+$ and $d_-$ are nonzero, we find the partially polarized solution 
		\be \label{partially_polarized}\bfd_{pp} = \frac1{\sqrt2} \bpm e^{i(\phi+\g)}\sqrt{\z B /v - \wt r/\wt u}, & 0,& e^{i(\phi-\g)} \sqrt{-\z B/v-\wt r/\wt u} \epm,\ee 
		which is only a solution provided that the field is less than the critical value $B_c$, where 
		\be B_c = \frac{v|\wt r|}{\z \wt u},\ee
		which only implicitly determines $B_c$ due to the implicit $B^2$ dependence of the coefficients on the RHS. When $B=0$, $\bfd_{pp}$ is invariant under $SO(2)_\bfd$ spin rotations about $ \bfd$, but unlike the fully polarized case the only $SO(3)$ rotation which acts in the same way as a gauge transformation is a spin flip, which acts in the same way as a gauge rotation by $\pi$. The order parameter manifold when $B=0$ is therefore $\mcm = (U(1)_g \times SO(3)_s) / (SO(2)_{\bfd} \times \zt ) = (U(1) \times S^2)/\zt$. 
		Since $\pi_1([U(1) \times S^2]/\zt) = \zz$, the $B=0$ phase has vortices and a nonzero $T_{BKT}$ \cite{mukerjee2006topological}. When $B\neq 0$, the order parameter manifold is instead $(U(1)_g \times U(1)_s)/\zt$.

		When $v<0$ the fully polarized solution is favored regardless of $B$, while when $v>0$ the partially polarized solution is favored as long as $B<B_c$. The transition at $B_c$ occurs when vortices in the spin mode $\gamma$ condense, and is of BKT type within the context of the present discussion. Since we know from experiments that the low-field phase is not fully polarized, we may therefore assume $v>0$ in what follows.

		\begin{figure}
			\subfloat[]{%
				\includegraphics[width=.33\textwidth]{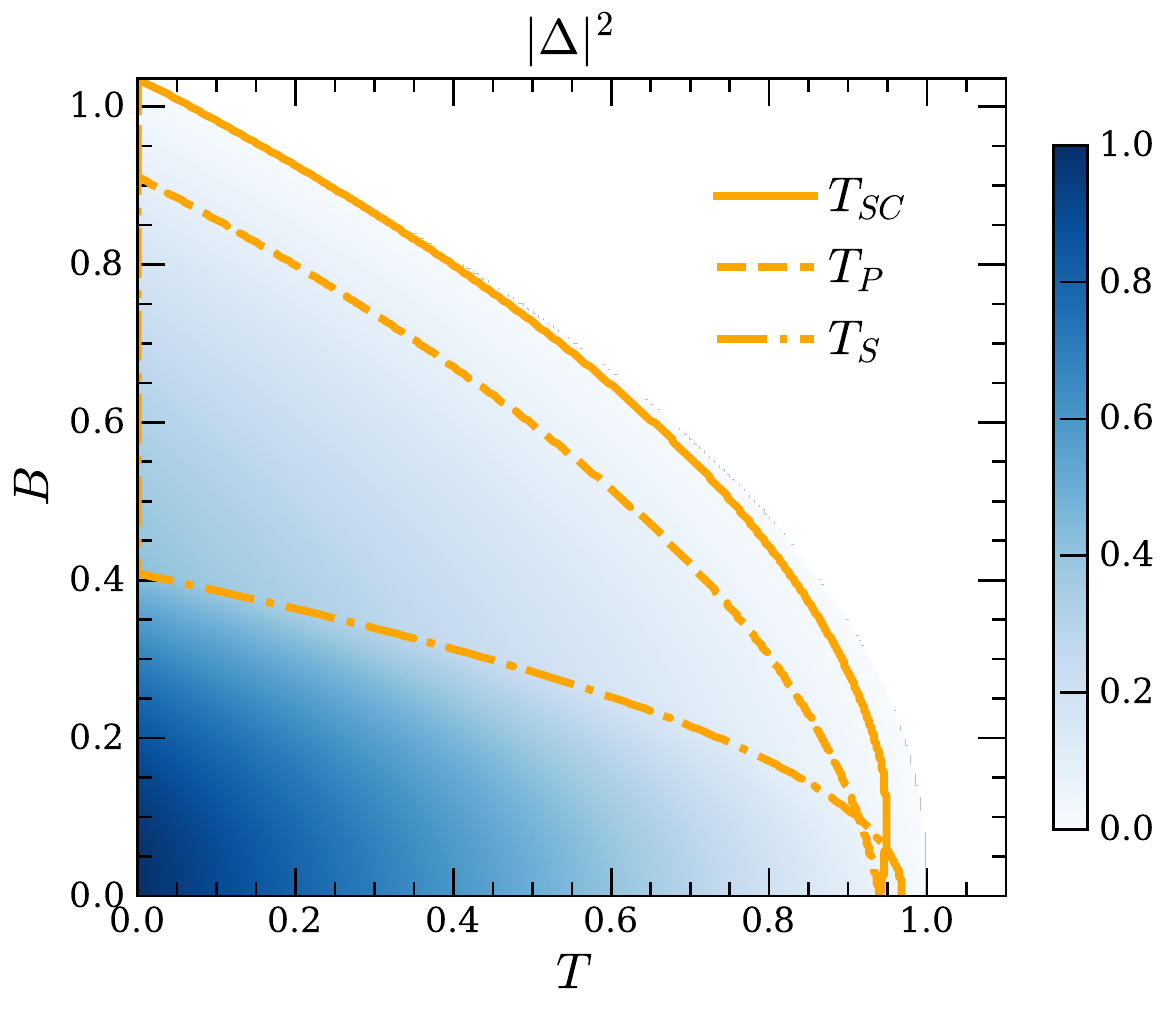}%
			}\hfill
			\subfloat[]{%
				\includegraphics[width=.33\textwidth]{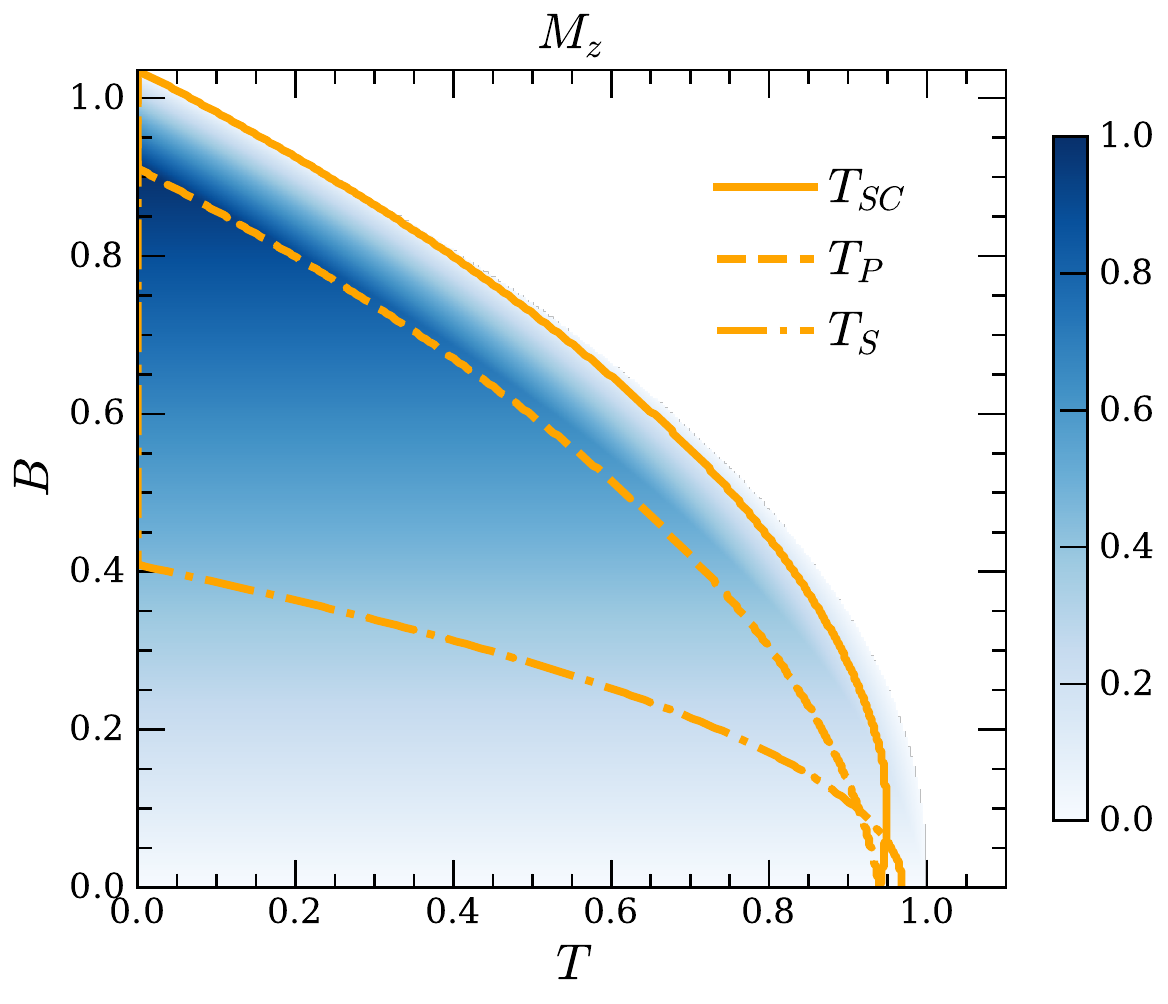}%
			}
			\hfill 
			\subfloat[]{%
				\includegraphics[width=.33\textwidth]{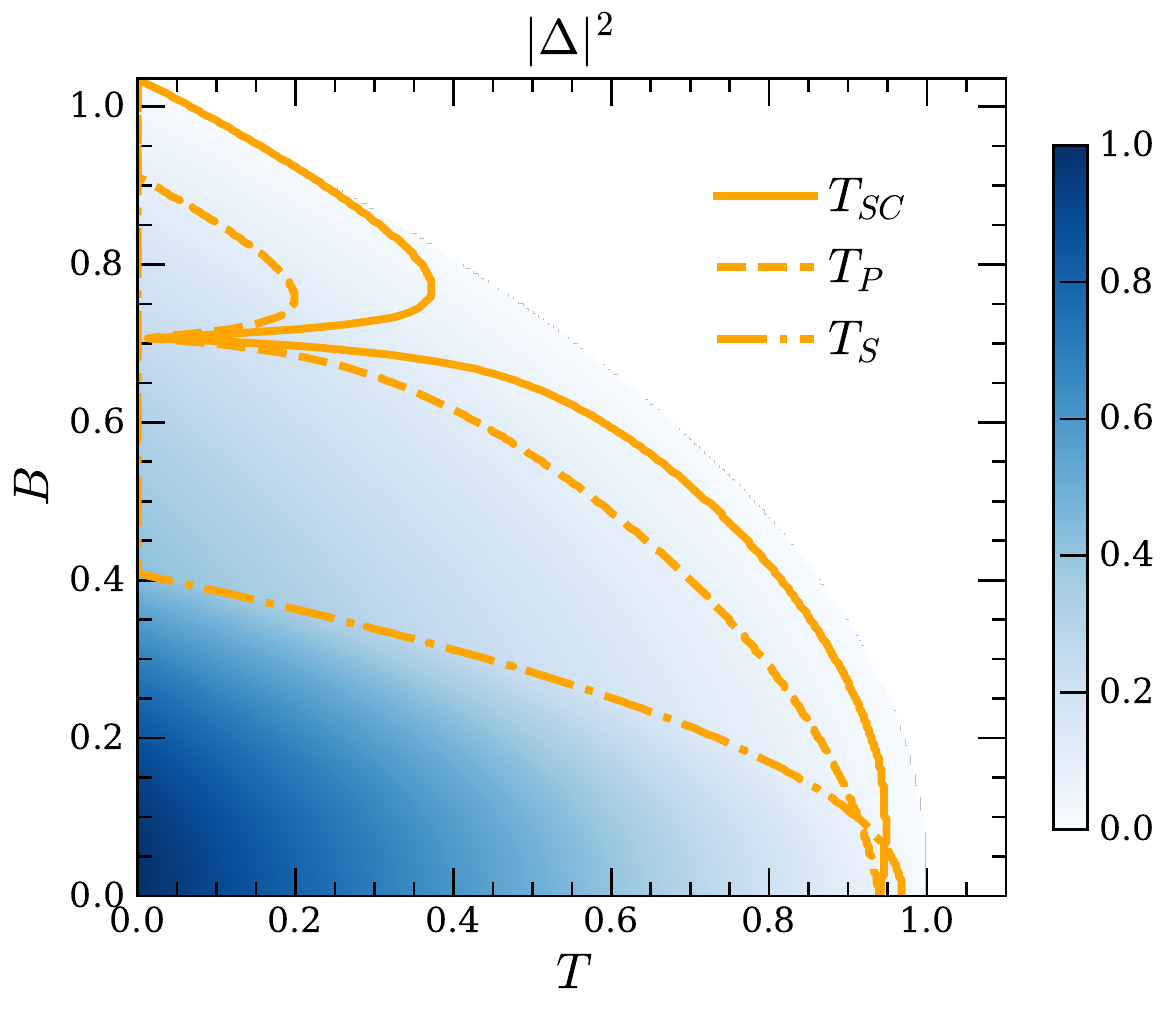}%
			}
			\caption{a) condensate fraction with no spin-valley locking (normalized by $|\De(B=0,T=0)|^2$) for a typical choice of parameters, and ignoring dependence of $K$ on $B$. $T_{S}$ denotes the transition temperature when the singlet component of the order parameter becomes zero, the line $T_P$ denotes the transition from the partially to fully polarized state, and $T_{SC}$ marks the superconducting transition temperature. b) expected value of the component of the magnetization $\bfM = \bfd_1\times\bfd_2$ along the field direction, normalized to $M_z(B=0,T=0)$. c) the same as a), but taking into account a downward renormalization of $K$ by a term proportional to $B^2$, leading to a field-induced Lifshitz transition. These types of phase diagrams are obtained by taking $-r,-r',u,u',\kappa,\kappa'>0$, $v,q\geq 0$, and $\z$ small and nonzero.}
			\label{fig:cond_fracs}
		\end{figure}
		
		Figure \ref{fig:cond_fracs} shows the condensate fraction $\lan |\De|^2\ran$ and magnetization $\lan M_z\ran$ for a typical choice of parameters. Panels a) and b) show the condensate fraction and magnetization, with the lines drawn at the locations of the BKT transitions separating the various phases. Panel c) shows where the transitions occur upon accounting for a further suppression of the stiffness $K$ by orbital effects (see appendix \ref{app:lg_coeffs}), with the Lifshitz transition occurring at the field where $T_{SC}$ first goes to zero. 
		
		Note that for any nonzero magnetic field, the transition into the fully polarized state always occurs before superconductivity is killed altogether ($T_P < T_{SC}$ in figure \ref{fig:cond_fracs}). This is because the fully polarized state is favored by a term which is linear in $|\De|^2$ (the $\bfB\cdot (\bfd_1\times\bfd_2)$ term), while the partially polarized state is favored by a term quadratic in $|\De|^2$ (the $v$ term); hence when $|\De|^2$ is small the fully polarized solution always wins. This means that in the $w=0$ limit, near the boundary of the superconducting dome the superconductor will always be in the fully polarized phase, regardless of the strength of the magnetic field. As a result, $T_{SC}$ {\it increases} linearly with $B$ at small $B$, by an amount which is small in $\z$. Since in the context of magic-angle TTG we expect $\z$ to be small, especially so at small $U$ (again, see appendix \ref{app:lg_coeffs}), this may or may not be incompatible with experiment.

		\section{Nature of the Lifshitz critical points \label{app:lifshitz}}
		
		In this appendix we elaborate on the nature of the Lifshitz critical points discussed in section \ref{sec:lifshitz}. Throughout we will write down free energies which involve only the phase mode $\phi$, with the fluctuations in the spin part of the order parameter ignored on grounds that they flow to zero coupling in the IR. 
		
		\sss{Isotropic case}
		
		We first review the isotropic case, where the Lagrangian at the $T=0$ critical point contains no quadratic derivative terms: 
		\be \label{iso_lifshitz} \mcl = \frac12\( (\p_\tau\phi)^2 + \vs (\D^2 \phi)^2\).\ee 
		A complete discussion of the finite-temperature physics of this model can be found in \cite{ghaemi2005finite}; here we will review the results of \cite{ghaemi2005finite} which are relevant for the present discussion. 
		
		The spatial correlator $\lan \phi(x)\phi(0)\ran$ calculated from \eqref{iso_lifshitz} is logarithmically divergent in the IR at $T=0$, and the model has QLRO. At nonzero $T$ the divergence is power law, so that the correlation functions of $e^{i\phi}$ are ultra short-ranged. 
		This latter fact does not mean that the model is necessarily disordered at all $T>0$ however, since as stressed in \cite{ghaemi2005finite} irrelevant terms will generate a nonzero stiffness (i.e, a nonzero $(\D \phi)^2$ term), allowing the $e^{i\phi}$ correlators to decay as power laws at $T>0$. 
		
		Accounting for these effects at $T>0$ is done by starting instead with the Lagrangian 
		\be \label{iso_renorm_l}\mcl = \frac12\( (\p_\tau\phi)^2 + K^0 (\D \phi)^2 + \vs (\D^2\phi)^2 + \eta (\D\phi)^4 \),\ee 
		where the marginally irrelevant $\eta$ term is responsible for generating a nonzero stiffness at $T>0$, and $K^0$ is a counterterm ensuring that the renormalized stiffness $K$ vanish at $T=0$. 
		
		The renormalized stiffness $K$ can be computed within the framework of an RPA (large-$N$) approximation. In this approximation, $K$ is given by the bare value $K^0$ in addition to a contribution coming from a geometric sum of 1-loop diagrams involving the $\eta$ interaction. Writing the renormalized $\phi$ propagator as $(\o^2 +  Kk^2 + \vs k^4)\inv$, we have 
		\bea K & = K^0 + 4\eta T \isum \frac{k^2}{\o_n^2 + K k^2 + \vs k^4}\\ 
		& =2 \eta\int_\bfk k^2\(\frac1{\sqrt{ K k^2+ \vs k^4}}- \frac1{\sqrt{\vs} k^2} + 2\frac1{\sqrt{ K k^2+ \vs k^4}} \frac1{e^{\sqrt{Kk^2+\vs k^4}/T}-1} \)  \\ 
		\eea 
		where we have used the requirement that $K_y=0$ at $T=0$ to solve for $K_y^0$. 
		
		Performing this integral and solving for $K$ at low $T$ gives \cite{ghaemi2005finite}
		\be K \approx 4 T \sqrt\vs \frac{\ln^2(\L^2/T)}{\ln(\L^2/T)},\ee 
		so that $K/T\sqrt\vs\ll1$ at low $T$.
		
		The value of $K$ determines $T_{BKT}$ by way of the Kosterlitz-Thouless criterion $K\pi/2 = T$. Since in the present case $K/T \ra 0$ at low $T$, there will {\it not} be a BKT transition at finite $T$, and the theory is disordered for all non-zero temperatures \cite{ghaemi2005finite}.
		
		\sss{Anisotropic case} 
		
		Let us now apply a similar analysis to the anisotropic case, where only one component of the stiffness vanishes at the $T=0$ critical point. 
		At $T=0$, we may work with the Lagrangian 
		\be\label{aniso_l} \mcl = \frac\l2\Big( (\p_\tau\phi)^2 +  K_x(\p_x\phi)^2 + \vs (\p_y^2\phi)^2\Big),\ee
		where $\vs$ is needed for stability. We take $\phi$ to be dimensionless, with the dimensionful factor of $\l$ out front having dimensions of $\p_y$. 
		At zero temperature the $\phi$ correlator along e.g. the $\uvx$ direction goes as 
		\bea \lan \phi(x) \phi(0)\ran & = \frac1\l \int_{\o,\bfk} \frac{e^{ik_xx}} {\o^2 + K_x k_x^2 + \vs k_y^4} 
		= \frac1\l\frac{2\G(5/4)^2}{\pi^{3/2}(\vs K_x)^{1/4}} \int \frac{dk_x}\twp \frac{e^{ik_xx}}{\sqrt{k_x}},\eea 
		which is IR finite. The theory therefore orders at $T=0$, unlike in the isotropic case \cite{moat}. However, like the isotropic case, the finite-$T$ correlator calculated using the above Lagrangian \eqref{aniso_l} has a power-law divergence in the system size: for example, at large $T$, 
		\be \lan \phi(x)\phi(0)\ran = \frac{T}{2^{5/2} K_x^{3/4}\l}  \int \frac{dk_x}\twp \frac{e^{ik_x x}}{k_x^{3/2}},\ee 
		which diverges as $\sqrt{L}$, with $L$ the system size. 
		
		As in the isotropic case, this divergence does not rule out the existence of a finite-$T$ phase with QLRO, as we must account for the presence of other terms which generate a nonzero $y$ component of the stiffness at $T>0$.  
		
		At $T>0$, the correct starting point is therefore instead, analogously to \eqref{iso_renorm_l},
		\bea \mcl = \frac\l2\Big( (\p_\tau\phi)^2 +  K_x(\p_x\phi)^2 + K_y^0 (\p_y\phi)^2 + \vs (\p_y^2\phi)^2 + \eta (\p_y\phi)^4\Big).\eea 
		
		The renormalized $y$ stiffness $K_y$ can be computed using the same 1-loop diagram as before.
		We have 
		\bea K_y & = K_y^0 + \frac{4\eta T}\l \isum \frac{k_y^2}{\o_n^2 + K_xk_x^2 + K_y k_y^2 + \vs k_y^4}\\ 
		& = \frac{2\eta}{\l} \int_\bfk k_y^2  \Bigg( \frac{1}{\sqrt{ K_xk_x^2 + K_y k_y^2+ \vs k_y^4}} - \frac{1}{\sqrt{ K_xk_x^2 + \vs k_y^4}} + 2\frac1{\sqrt{ K_xk_x^2 + K_y k_y^2+ \vs k_y^4}} \frac1{e^{\sqrt{ K_xk_x^2 + K_y k_y^2+ \vs k_y^4}/T}-1}\Bigg) .
		\eea

		Assuming $K_y \ll \sqrt\vs T$, we can ignore the $K_y$ dependence of the third term in the integral. We then solve for $K_y$, finding 
		\be K_y \approx \vup T^{3/2},\qq \vup = \frac{C}{\vs^{3/4}\sqrt{K_x} \( 2\pi^2 \l/\eta + 2\L/\sqrt{K_x}\vs\)}\ee 
		where $C\approx 11$. 
		At small $T$ we indeed have $K_y \ll \sqrt\vs T$, justifying our earlier assumption. Note that unlike in the isotropic case, here we are not in the upper critical dimension of the $T=0$ theory, and the term $\eta$ which generates $K_y$ is not marginal. Therefore there is no logarithmic dependence of $K_y$ on $T$, and the dependence of $K_y$ on the value of $\eta$ can not a priori be neglected at low $T$. 
		
		In the anisotropic case, the stiffness which determines the location of a possible BKT transition is set by $K_{\rm eff} = \l \sqrt{K_xK_y}$. Since $K_{\rm eff} \propto T^{3/4}$, $K_{\rm eff}$ will always be larger than $2T/\pi$ at low enough $T$, meaning that there will always be a finite-temperature BKT transition. Explicitly, in the present approximation the transition occurs at 
		\be T_{BKT} \approx \(\frac{\l\pi}{2} \sqrt{K_x \vup}\)^4.\ee 
		The value of $T_{BKT}$ is of course non-universal, and depending on the microscopics may very well be small enough to be essentially indistinguishable in experiments from zero. 
		
		\sss{Critical currents} 
		
		In this section we compute the critical current in the region of parameter space near the Lifshitz transition. 
		
		First consider the normal phase where the pairing occurs at zero momentum. In this case we may examine the free energy density 
		\be f[\D\phi] = \frac\r2\(\g_x (\p_x\phi)^2 + \g_y(\p_y\phi)^2\) + r \r +\frac{u}2\r^2,\ee 
		where $\g_x,\g_y,u,-r>0$. To determine the critical current, we follow the standard procedure \cite{tinkham2004introduction} of Legendre transforming to a thermodynamic potential $g$ which is a function of the current $\bfI$, satisfying (working in units where $2e=1$)
		\be g[\bfI] = -\D \phi \cdot \bfI + f[\D\phi],\qq \frac{\d g}{\d \bfI} = -\D\phi ,\qq \frac{\d f}{\d \D\phi } = \bfI.\ee 
		Therefore 
		\be g[\bfI] = -\frac{I_x^2}{2\r\g_x} - \frac{I_y^2}{2\r\g_y} + r\r + \frac{u}2\r^2.\ee 
		Minimizing over $\r$ gives 
		\be \frac{I_x^2}{2\g_x} + \frac{I^2_y}{2\g_y} =  |r|\r^2 - u \r^3.\ee 
		Maximizing the RHS, we then see that the critical currents along the $x$ and $y$ directions are given by 
		\be I_{c,x} = \sqrt{\frac{8\g_x }{27} \frac{|r|^3}u},\qq I_{c,y} = \sqrt{\frac{\g_y}{\g_x}} I_{c,x}.\ee 
		
		Now we consider a situation which favors finite-momentum pairing along the $\uvy$ direction. We consider the free energy density
		\be f[\Psi] = \frac{1}{2}\( \g_x|\p_x \Psi|^2 - \g_y|\p_y\Psi|^2 + \frac\vk2 |\p_y^2\Psi|^2\) + r|\Psi|^2 + \frac u2|\Psi|^4, \ee 
		with $\g_y,\vk>0$. We then consider an FF state, with 
		\be \Psi = \sqrt{\r}e^{iq_0y + i\phi},\qq q_0 = \sqrt{\g_y/\vk},\ee 
		and assume that $\phi$ varies on scales much longer than $1/q_0$. We can then consider the approximate free energy 
		\be f[\D \phi] \approx \frac\r2\(\g_x (\p_x\phi)^2 + \frac{\g_y}2 (\p_y\phi)^2\) + (r-\g_y^2/4\vk)\r + \frac u2\r^2 .\ee 
		Repeating the same steps as above leads to critical currents of 
		\be I_{c,x} = \sqrt{\frac{8\g_x}{27u^2} |r -\g_y^2/4\vk|^3},\qq I_{c,y} = \sqrt{\frac{\g_y}{2\g_x}} I_{c,x}.\ee 
		
		Consider finally the scenario where both quadratic derivatives in the free energy density have negative coefficients \cite{buzdin2007fflo,samokhin2017current}, with 
		\be f[\Psi] = \frac12\(-\g_x |\p_x\Psi|^2 - \g_y |\p_y\Psi|^2 + \frac{\vk_x}2|\p_x^2\Psi|^2 + \frac{\vk_y}2 |\p_y^2\Psi|^2 + \vk_{xy} |\p_x\p_y\Psi|^2 \) + r|\Psi|^2 + \frac u2 |\Psi|^4,\ee 
		with all coefficients except $r$ positive. 
		Consider again an FF state $\Psi = \sqrt\r e^{iq_0y + i\phi}$, with finite momentum in the $\uvy$ direction. Assuming $\g_y \vk_{xy}/\vk_y - \g_x \geq 0$ and making the same approximations as above, this leads to the critical currents 
		\be I_{c,x} = \sqrt{\frac{8(\g_y\vk_{xy}/\vk_y - \g_x)}{27u^2} |r-\g_y^2/4\vk_y|^3},\qq I_{c,y} = \sqrt{\frac{\g_y}{2(\g_y \vk_{xy}/\vk_y - \g_x)}}I_{c,x}.\ee

		\section{Estimate of the in-plane critical field $B_{c}$ \label{app:best}}

		In this appendix we provide a crude estimate of the critical in-plane field $B_{c}$ of the superconductor at zero displacement field and optimal doping. Given the phenomenological nature of our analysis and the fact that $B_{c}$ may very well not be small enough for leading-order perturbation theory in $B$ to be reliable, this calculation is not meant to be taken as a sharp prediction --- rather it is meant simply as a consistency check to make sure that our proposal is not grossly incompatible with experiment. 
		
		From equation \eqref{renorms} of appendix A, the stiffness at zero displacement field and small $B$ is given by\footnote{Note that we are ignoring contributions to the stiffness which arise from Zeeman effects. Within the framework of appendix A, Zeeman effects in the mirror-even sector do not contribute to the stiffness at leading order, while contributions from the mirror-odd sector are small (being suppressed by a factor of $\l (\mu_B/v_D\d)^2 \ll 1$), and will be ignored.} (assuming for simplicity that the velocity of the mirror-odd Dirac cone after integrating out the mirror-even fermions is isotropic and equal to the monolayer value of $v_D$)
		\be K_{eff} = K_0 - 2C_2|\De|^2\l h v_D^2 \d^2 B^2 + \cdots ,\ee 
		where $\cdots$ are terms higher order in $B$, $\d \approx 0.3$nm is the interlayer separation, $K_0$ is the $B=0$ phase stiffness, and where the definitions of $C_2,\l,h$ will be recapitulated shortly. Estimating the zero-field stiffness as $K_0\approx 2T_{BKT,0}/\pi$ where $T_{BKT,0}\approx 1.5$K \cite{inplane_fields} is the zero-field transition temperature, we have 
		\be B_c \approx \frac1{v_D\d} \sqrt{\frac{ T_{BKT,0}}{\pi C_2 |\De|^2 \l h }} \approx 1.2  \, \sqrt{\frac{\mce}{1\, {\rm mev}}}\, {\rm T},\ee 
		where the energy scale $\mce$ is defined as 
		\be \mce \equiv \frac1{\pi C_2|\De|^2\l h}.\ee 
		We now need to estimate the quantities appearing in $\mce$. 
		
		The coupling $h$ is associated with the pairing terms for the mirror-odd fermions that arise upon integrating out the mirror-even sector (see \eqref{hdef}), and has dimension of inverse energy squared. While we are not in a position to calculate $h$ exactly, on general grounds we expect $h \sim 1/\ep_{F}^2$, where $\ep_F$ is the Fermi energy of the mirror-even bands which become superconducting. $\ep_F$ is not known exactly and its determination depends on knowledge of what happens to the correlated insulator at $\nu=-2$ upon hole doping, but a value in the ballpark of a few tens of mev seems appropriate. 
		
		$C_2$ can be read off from an expansion of \eqref{second_o_matsu}, giving 
		\be C_2 = -\frac3T \int_{\bfk,x} y \,\sech^2(k/2 T) \frac{k - T \sinh(k/T)}{2 k^3} = \frac1{8\pi T},\ee 
		where $x$ is integrated from 0 to 1 and $y=x(1-x)$. 
		For the present purposes of estimating $B_c$, we will take $T \ra T_{BKT,0} \sim T_{onset}$ in the above. 
		
		The appropriate value to take for $\De$ (equal to the superconducting gap in our mean-field treatment) is not currently known experimentally. We will leave $\De$ as a free parameter for now, and note only that due to the strong-coupling nature of the problem and the energy scales involved, a value of $\De\sim 1$ mev does not seem unreasonable. 
		
		Finally, we need to estimate the dimensionless parameter $\l$, which appears in our Lagrangian as $\l \psi_- \De^\da \psi_- + h.c.$ and originates from a mean-field decoupling of interactions of the form $\psi_-^\da \psi_-^\da \psi_+ \psi_+ + h.c.$ (all indices suppressed). These interaction only conserve the number of mirror-even and mirror-odd fermions modulo 2, and they do not appear directly in the regular density-density interaction, as the total fermion density is $\d \r = \d\r_{even} + \d\r_{odd}$ which separately conserves the number of mirror-even and mirror-odd fermions. These interactions furthermore do not arise at first order upon projecting the density-density interaction to the active bands, since at $D=0,B=0$ the form factors of the active bands do not mix the mirror-even and mirror-odd sectors (by symmetry). Therefore after projecting the interaction to the active bands, $\l$ can only arise from an interaction corresponding to a second-order process involving tunneling to remote bands. The strength of these higher-order interactions is roughly $U^2/\De E$, where $\De E$ is the gap to the remote bands and $U$ is the Coulomb interaction strength; $\l$ is accordingly $\l \sim (U^2/\De E)/U = U/\De E$. Estimating $U \sim e^2/(4\pi \ep_0 \ep_r \sqrt{A_{moire}}) \sim 30$ mev (with the relative dielectric constant $\ep_r \sim 5$) and $\De E \sim 50$ mev (based on a naive reading of the band structure \cite{khalaf2019magic,fischer2021unconventional} gives $\l \sim 3/5$. 
		
		With the above estimates in hand, we may plug in numbers and write 
		\be \label{best} B_c \approx 28\,\, \frac{1\, {\rm mev}}{|\De|} \frac{\ep_F}{20 \, {\rm mev}} \sqrt{\frac{3/5}{\l}} \sqrt{\frac{C_2\inv}{8\pi T_{BKT,0}}}\,\, {\rm T}.\ee 
		The value of $B_c$ at zero displacement field and optimal doping is experimentally determined to be around 12 T \cite{inplane_fields}. This is rather close to the value we would read off from \eqref{best} given the estimates for the parameters on the RHS as discussed above, and given the crudeness of the present estimate (especially in the determination of e.g. $h$ and $C_2$) this is really the best we may hope to do. A more careful analysis is left to future work, but we find it encouraging that the rough estimate here at least seems to be in the right ballpark.

		\section{Single-particle physics \label{app:single_particle}} 
		
		In this appendix we provide a few details on  the single-particle properties of TTG in the presence of an in-plane field. 
		
		\ss{Hamiltonian and symmetries} 
		
		In the basis $(\psi_1,\psi_2,\psi_3)$, with $\psi_l$ the annihilation operator for fermions on layer $l$, the single-particle Hamiltonian is 
		\be \label{appham}  H_0 = \bpm -iv_D\bfsig_{v,\t/2} \cdot (\D-i \bfA_1) + U/2 &  T &\\ T^\da & -iv_D\bfsig_{v,-\t/2}\cdot(\D-i \bfA_2) & T^\da \\ 
		&T & -iv_D\bfsig_{v,\t/2}\cdot(\D-i\bfA_3) -U/2 \epm  + H_Z,\ee 
		where $\bfsig_{v,\t/2}$ denotes a rotation of $\bfsig_v = (\tau^z \s^x,\s^y)$ by $\t/2$, where $\t$ is the twist angle. $\bfA_l$ and $H_Z$ are as in the main text. 
		
		We will use the form of the $T$ matrix given in e.g. \cite{tarnopolsky2019origin}, which we take to be  
		\be T(\bfx)
		= \sum_{i=1}^3 e^{-i \tau^z\bfq_i \cdot \bfx} T_i,\ee 
		with the tunneling matrices 
		\be T_i =  w_{AA} \unit + w_{AB} (\s^x\cos[\twp (i-1)/3]+ \tau^z \s^y \sin[\twp (i-1)/3] )\ee 
		and where 
		\be \bfq_1 = k_\t (0,-1),\qquad \bfq_2 = k_\t (\sqrt3/2,1/2),\qquad \bfq_3 = k_\t (-\sqrt3/2,1/2),\ee 
		with 
		\be k_\t \equiv 2k_D \sin(\t/2),\qq  k_D \equiv \frac{4 \pi}{3a_0},\ee 
		and with $a_0 \approx 0.246$nm the graphene lattice constant. We will work in conventions where the Moire $K$ and $K'$ points are given as $K = -\bfq_2$ and $K' = \bfq_3$. In the limit where the graphene layers are decoupled, the band structure then consists of two Dirac cones at $K$, and one at $K'$.

		In the limit $w_{AA}=0$ the rotations of the $\bfsig_v$ matrices in \eqref{appham} can removed by a unitary transformation \cite{tarnopolsky2019origin}, but even away from this limit their effects can usually be ignored on the grounds that $\t$ is small \cite{tarnopolsky2019origin,bistritzer2011moire}. We will therefore simply ignore the rotations in what follows. 
		
		The symmetries of the $U=\bfB=0$ problem include \cite{khalaf2019magic,cualuguaru2021tstg}\footnote{$C_{2x}$ is technically only a symmetry in the limit where the $\pm \t/2$ rotations of $\bfsig_v$ can be ignored. }
		\bea C_{2z} & = \s^x\tau^x R_\pi \\ 
		C_3 & = e^{\twp i \s^z\tau^z/3} R_{2\pi/3}\\
		C_{2x} & = M \s^x R_y\\ 
		\mct & = \tau^x\mck,
		\eea 
		where the RHSs are the matrices which appear in the adjoint action on the $\psi_l$ fields and $\mck$ is complex conjugation. 
		Here $R_\phi$ is a rotation through $\phi$ acting in the $xy$ plane, $R_y$ implements a reflection of the $y$ coordinate, and the mirror symmetry $M$ acts in layer space as 
		\be M = \bpm &&1 \\ & 1 & \\ 1&&\epm.\ee 
		When written in terms of fields which diagonalize $M$, the above expression for $H_0$ in \eqref{appham} reduces to the form given in the main text \eqref{h0}.
		
		As in the bilayer case, in the $w_{AA}=U=0$ limit we have $\{ \s^z,H\}=0$. Unlike in TBG, away from this limit there is no unitary chiral symmetry which arises if we ignore the $\pm \t/2$ rotations of $\bfsig_v$ (which in TBG acts as $i\s^x \mu^y \mck$, with $\mu^y$ in layer space). 
		
		Turning on a nonzero $U$ breaks $M$ and $C_{2x}$, and turning on a nonzero $\bfB$ breaks $C_3,\mct,M,$ and $C_{2z}$, and breaks $C_{2x}$ unless $\bfB\prl \uvx$. Any direction of field preserves $C_{2z}M, C_{2z}\mct,$ and $M\mct$. The combination of $C_{2x}$ and one of $\mct,M,C_{2z}$ is also preserved if $\bfB \prl \uvy$. Like in the bilayer case, $C_{2z}\mct$ is unbroken for any choices of $U,\bfB$, and prevents the innermost bands from separating from one another \cite{po2018origin}.

		\ss{Band structure} 
		
		\begin{figure}
			\includegraphics[width=.95\textwidth]{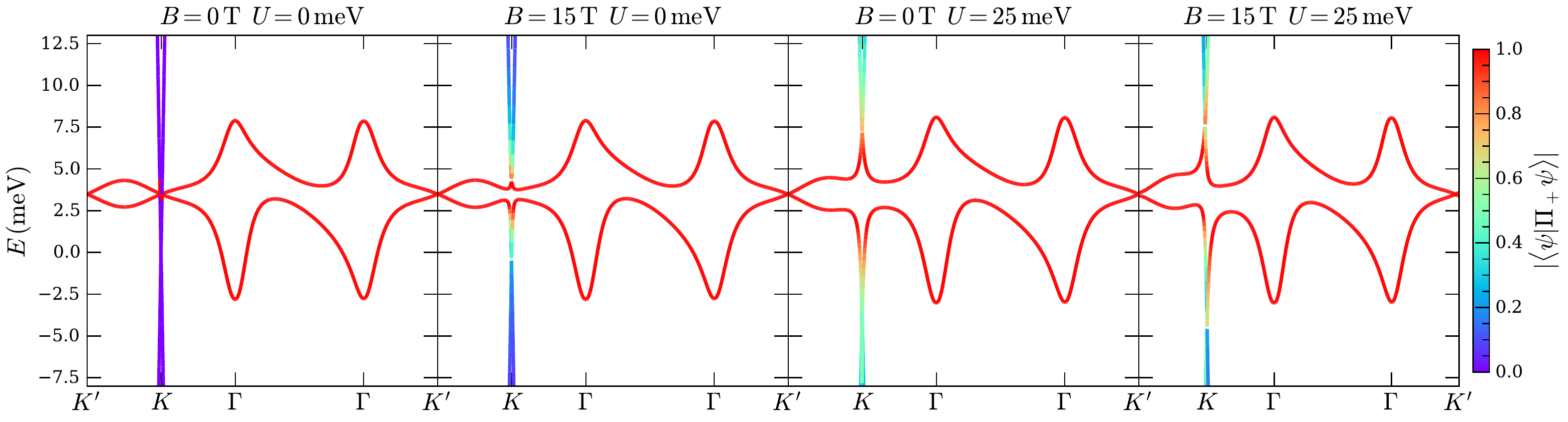}
			\caption{\label{fig:band_0tr} The structure of the central bands for several values of $U$ and $B$ (here $\bfB\prl\uvx$). The color scale denotes the extent to which the wavefunctions are mirror-even, with $\Pi_+ = (1 +M)/2$. }
		\end{figure} 
		
		We now briefly discuss the band structure obtained from \eqref{appham}. For concreteness we will fix $\t = 1.57^\circ, w_{AB} = 110$meV, and $w_{AA} = 0.8 w_{AB}$. Since we are mainly interested in the orbital effects of the field, we will ignore the Zeeman energy. 
		
		Similarly to $U$, the main orbital effects of $\bfB$ on the band structure are to hybridize the mirror-odd Dirac fermion $\psi_-$ with the mirror-even flat bands. Examples of the band structure in the graphene $K$ valley for various values of $U,\bfB \prl \uvx$ are shown in figure \ref{fig:band_0tr}, where the color denotes the magnitude of the mirror-even component of the band wavefunctions. 
		
		The orbital effects of the magnetic field are rather different than in the case of TBG \cite{kwan2020twisted}. First, when $U=0$, mirror symmetry implies that the spectrum at each $\bfk$ point is an {even} function of $B$. This means that at small $U$ and small $\bfB$, the orbital effects on the band structure are rather weak. Secondly, despite the explicit breaking of $C_3$ by the magnetic field, Dirac points remain at the Moire $K,K'$ points to an extremely good approximation for all experimentally relevant field strengths. 
		


		\ss{Effects of the field near the $K$ and $K'$ points}
		
		In this section we adopt the approach used in \cite{bistritzer2011moire} to make some simple statements about the effects of the field on the band structure near the Moire $K$ and $K'$ points, at small values of $U,B$. Throughout we will focus on the graphene $K$ valley, and will neglect the physical spin degree of freedom. We will also find it convenient to follow \cite{tarnopolsky2019origin} and use the notation $\a \equiv w_{AB} / v_Dk_\t$ and $\kappa \equiv w_{AA}/w_{AB}$. 
		
		\sss{$K'$ point} 
		
		We can understand the (lack of) effects of the field at the $K'$ point in a rather pedestrian manner as follows. We follow \cite{bistritzer2011moire} and introduce a minimal model for the modes right at $K'$ by writing down the Hamiltonian 
		\be H_{K'} = \bpm 0 & T_1 & T_2 & T_3 & T_1^\da & T_2^\da & T_3^\da \\ 
		T_1^\da & h_1 &&&&&\\
		T_2^\da &&h_2 &&&& \\
		T_3^\da &&&h_3 &&&  \\ 
		T_1 &&&&-h_1 &&\\
		T_2 &&&&&-h_2& \\
		T_3 &&&&&&-h_3 \epm, \ee 
		where we are in the basis $(\psi_2,\psi^1_1,\psi^2_1,\psi_1^3,\psi_3^1,\psi_3^2,\psi_3^3)^T$, with the lower index denoting the layer index, the upper index denoting a reciprocal space lattice index, and where (working in units where $v_D=1$)
		\be h_i = (\bfq_i - \bfA) \cdot \bfsig + U/2.\ee
		As in the analysis of the analogous bilayer problem, there are two zero modes in this simplified 14-band problem. Let $\psi^\upa_2 = (1,0)^T$ and $\psi^\doa_2 = (0,1)^T$. We will then have a zero mode  provided we take 
		\be \psi^j_1 = -h_j\inv T^\da_j \psi^2_\s,\qquad \psi_3^j = +h_j\inv T_j \psi^2_\s,\ee 
		and that 
		\be \sum_j ( T_j h_j\inv T^\da_j - T^\da_j h_j\inv T_j)\psi_2^\s=0,\ee 
		with this last equation trivially satisfied as our $T_j$ matrices are Hermitian. 
		
		For simplicity we will now work perturbatively in $\bfA$. At $\bfA=0$, the zero modes $\k{\psi_{zm;\s}}$ have norm 
		\bea \lan \psi_{zm;\s}| \psi_{zm;\s}\ran &= 1 + 2 \sum_j T_j (h_j\inv )^2 T_j \\ 
		& = 1 + 6 \a^2 \vpi (\kappa^2+1),\qq \vpi \equiv \frac{1+(U/k_\t)^2/4}{(1-(U/k_\t)^2/4)^2}.\eea  
		After correctly normalizing the wavefunctions, the effective Hamiltonian in the zero-mode subspace near the $K'$ point then reads 
		\bea H_{eff,K'+\bfk}  & = \frac1{1 + 6 \a^2 \vpi (\kappa^2+1)} \(\bfsig\cdot\bfk + \sum_j (T_jh\inv_j  \bfsig\cdot(\bfk-\bfA) h\inv_jT_j + T_jh_j\inv \bfsig\cdot(\bfk +\bfA) h_j\inv T_j)  \)\\ 
		& = \frac{1-6\a^2}{1 + 6 \a^2\vpi(\kappa^2+1)} \bfsig \cdot\bfk,\eea 
		with the Fermi velocity out front confirming that when $U=0$, $\t_1^3 = \sqrt2 \t_1^2$, with $\t_i^l$ the $i$th magic angle for twisted $l$-layer graphene \cite{tarnopolsky2019origin,bistritzer2011moire}.
		As expected from symmetry considerations, $H_{eff,K'+\bfk}$ is independent of the magnetic field to this order. 
		
		\sss{$K$ point} 
		
		The effects of the magnetic field near the $K$ point are slightly more interesting, due to the hybridization between the flat bands and the mirror-odd Dirac cone. 
		
		We proceed as above: near $K$, the analogous minimal model operates in the space $(\psi_1,\psi_3,\psi_2^1,\psi_2^2,\psi_2^3)$, where again the upper index is the layer index and the lower index is the reciprocal lattice index. Right at $K$, the Hamiltonian is 
		\be H_K = \bpm 0 && T_1^\da & T_2^\da& T_3^\da \\ 
		&0&T_1^\da & T_2^\da & T_3^\da \\ T_1 & T_1 & h_1 &&\\ 
		T_2 & T_2 &&h_2 & \\ 
		T_3 & T_3 &&&h_3 \epm \ee 
		There are now four zero modes. We write the first four components of each zero mode (in the above basis) as the vector $(\psi^\s_1,\psi^\r_3)$ ($\s,\r = \upa/\doa$ as before). In order to have a zero mode, the $\psi_2^i$ components must satisfy 
		\be  \psi^i_2 = - h_i\inv T_i(\psi^\s_1 + \psi^\r_3)\ee 
		and 
		\be \sum_i T_i^\da \psi_i^2 = -\sum_i T_i^\da h\inv_i T_i(\psi^\s_1 + \psi^\r_3) = \frac{3\a^2U}2\frac{1 +  \kappa^2}{1 - (U/k_\t)^2/4} (\psi^\s_1 + \psi^\r_3) = 0.\ee
		Evidently we only have zero modes if $U=0$ (which we already know to be the case by looking at the band structure). 
		
		To isolate the effects of the magnetic field on the Dirac cones, we will accordingly now set $U=0$. Treating $\bfk$ and $\bfA$ as perturbations, the effective Hamiltonian in the zero mode subspace works out to be 
		\bea  H_{eff,K+\bfk}& = \frac1{1+3\a^2(1+\kappa^2)} \bpm \bfsig\cdot(\bfk(1-3\a^2) + \bfA) & -3\a^2 \bfsig\cdot\bfk \\ -3\a^2 \bfsig\cdot\bfk &  \bfsig\cdot(\bfk(1-3\a^2) - \bfA) \epm   \eea 
		When $\bfk=0$, we solve for the spectrum and find two doubly degenerate states at energies (as before, $\d$ is the interlayer separation)
		\be \ep_K = \pm \d |B|.\ee  
		These doubly degenerate states form Dirac cones which live at the $K$ point above / below zero energy.
		
		\begin{figure}
			\includegraphics[width=.65\textwidth]{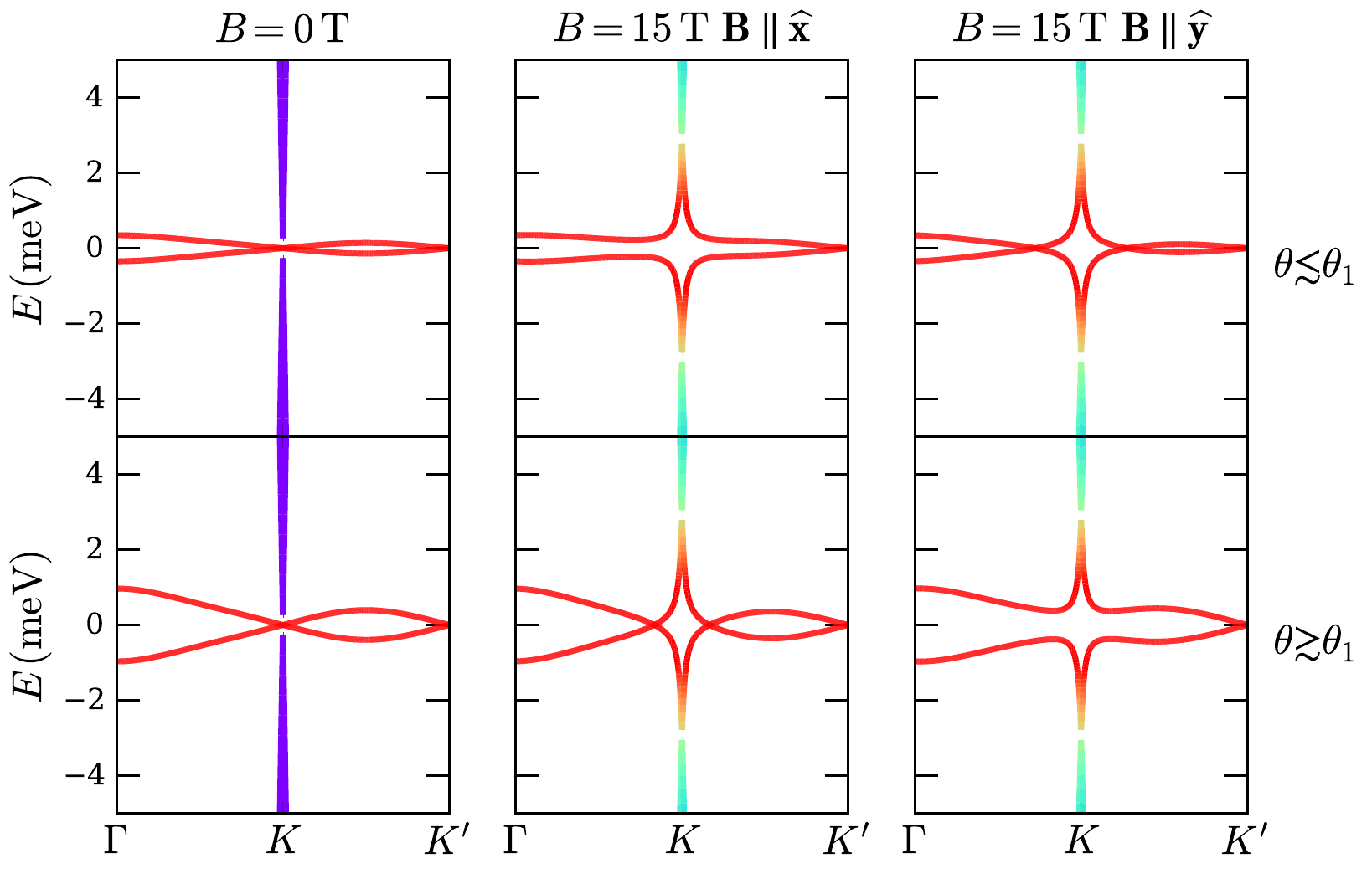}
			\caption{\label{fig:more_bandstr} Field effects of the band structure near the $K$ point, shown in the limit $w_{AA} = U = 0$. The first row of plots are at angles slightly less than the magic angle $\t_1$, while the plots in the second row are slightly above. For $\t<\t_1$ ($\t>\t_1$), two Dirac cones move off from $K$ along (normal to) the field direction (note that the momentum cut is along $\uvy$). The color coding indicates mirror projection, as in figure \ref{fig:band_0tr}.} 
		\end{figure}
		
		By computing the band structure numerically (see figure \ref{fig:more_bandstr}), one finds that when $\t\gtrsim\t_1$ (with $\t_1$ the first magic angle), two Dirac cones split off from the $K$ point and move along the direction normal to $\bfB$. To demonstrate this within the context of the present ten-band model we can consider e.g. $\bfA,\bfk \prl \uvx$, for which the effective Hamiltonian is 
		\be H_{eff,K+\bfk} \propto \s^x \tp \bpm k(1-3\a^2) + \d B & -3\a^2 k \\  -3\a^2 k & k(1-3\a^2) - \d B \epm\ee 
		The determinant of this matrix vanishes at $k=\pm k_*$, where 
		\be k_* = \sqrt{\frac{2}{1-6\a^2}}\d |B|,\ee 
		which evidently is only a solution provided that $1-6\a^2 > 0$, i.e. if $\a < 1/\sqrt{6}$, which in this approximation is the same as $\t > \t_1$. Therefore when $\t>\t_1$ we have two Dirac cones located at the $K$ point at energies $\pm \d B$, and two at zero energy at the momenta $(k_*,0)$. Since the orbital magnetic energy is much smaller than $k_\t$, this shift is rather small at experimentally relevant field strengths. 
		
		On the other hand, when $\t\lesssim\t_1$, one finds that the two Dirac cones which split off of the $K$ point move off parallel to $\wh\bfB$. To demonstrate this we may take $\bfk\prl \uvx, \bfA \prl \uvy$. The Hamiltonian is then 
		\be H_{eff,K+\bfk} \propto \s^x \tp \bpm k(1-3\a^2) & -3\a^2 k \\ -3\a^2 k & k(1-3\a^2) \epm + \d B \, \s^y \tp \s^z.\ee 
		Taking the determinant, one finds two doubly degenerate zero modes when $k=\pm k_*$, where now 
		\be k_* =  \frac{\d |B|}{\sqrt{6 \a^2 - 1}}.\ee 
		This is only a solution if $\a > 1/\sqrt{6}$, i.e if $\t < \t_1$. Note that the square root dependence of $k_*$ on $1/|1-6\a^2|$ and the linear dependence on $B$ holds on both sides of the magic angle.
		
		\bibliography{ttg} 
		\ms 
		\ms 
	\end{widetext} 
	
\end{document}